\documentclass[preprint,preprintnumbers,showkeys,amsmath,amssymb]{revtex4}

\usepackage{graphicx}
\usepackage{dcolumn}
\usepackage{bm}
\newcommand{\p}{\partial}
\newcommand{\be}{\begin{equation}}
\newcommand{\ee}{\end{equation}}
\newcommand{\mc}{$\mathcal{C}$}

\begin{document}


\title{Dynamics and stability of chimera states in two coupled populations
of oscillators}

\author{Carlo R. Laing}
 \email{c.r.laing@massey.ac.nz}
\affiliation{School of Natural and Computational Sciences, 
Massey University,
Private Bag 102-904 
North Shore Mail Centre, 
Auckland,
New Zealand
}%

\date{\today}

\begin{abstract}
We consider networks formed from two populations of identical
oscillators, with uniform strength all-to-all
coupling within populations, and also between populations, with a different strength.
Such systems are known to support chimera states in which oscillators within
one population are perfectly 
synchronised while in the other the oscillators are incoherent, and have a different mean
frequency from those in the synchronous population. Assuming that the oscillators in the incoherent
population always lie on a closed smooth curve \mc, we derive and analyse the
dynamics of the shape of \mc \ and the probability density on \mc, for four different types
of oscillators. We put some previously derived results on a rigorous footing, and analyse
two new systems.

\end{abstract}

\keywords{Chimera states, Coupled oscillators, Bifurcations, Collective behavior in networks}
\maketitle

\section{Introduction}
Chimera states in networks of coupled oscillators have been intensively studied in recent
years~\cite{panabr15,ome18}. Often they are studied in 
one-dimensional~\cite{omeome13,abrstr06,abrstr04,kurbat02} or
two-dimensional domains~\cite{lai17,omewol12,panabr15a,xiekno15,shikur04,marlai10} with nonlocal coupling,
but it was Abrams et al.~\cite{abrmir08} who first ``coarse-grained'' space and studied
chimeras in a network formed from two populations of oscillators, with equal strength
coupling between oscillators within a population, and weaker coupling to those in the other
population. Later studies of networks with such structure 
include~\cite{lai09a,precha17,marbic16,panabr16,pikros08} and we also mention the
experimental results~\cite{tinnko12,marthu13} and the prior
work~\cite{monkur04}. In such networks a chimera state occurs when one population
is perfectly synchronised (all oscillators behave identically) while in the other the oscillators
are not phase synchronised but all have the same time-averaged
frequency, which is different from that
of the synchronous population.
Such a state is similar to that of 
{\em self-consistent partial synchrony}~\cite{clupol18,vre96,clupol16}











Regarding the types of oscillators used, early works used phase 
oscillators with sinusoidal interaction functions~\cite{abrstr06,abrstr04},
while later studies include oscillators near a SNIC bifurcation~\cite{vulhiz14},
van der Pol oscillators~\cite{omezak15}, oscillators with inertia~\cite{boukan14,olm15,belbri16},
Stuart-Landau oscillators~\cite{lai10,precha17},
and neuron models including leaky integrate-and-fire~\cite{olmpol11}, quadratic 
integrate-and-fire~\cite{ratpyr17},
and FitzHugh-Nagumo~\cite{omeome13}.

The vast majority of papers concerning chimeras show just the results of numerical simulations
of finite networks of oscillators. Early researchers showed the existence of chimeras
using a self-consistency argument~\cite{abrstr04,kurbat02,shikur04,marlai10} and later the 
Ott/Antonsen ansatz~\cite{ottant08} was used to investigate their 
stability~\cite{lai09,abrmir08,marbic16,omewol12,mar10}.
However, these techniques relied on the number of oscillators being infinite, and more 
restrictively, that the oscillators were phase oscillators coupled through a purely
sinusoidal function of phase differences. (Also, states found using the Ott/Antonsen ansatz
are not attracting for networks of identical oscillators --- heterogeneity is required
to give stability~\cite{lai09,lai17}.) Finite networks of identical sinusoidally coupled
phase oscillators have been studied using the Watanabe/Strogatz 
ansatz~\cite{watstr94,pikros08,panabr16}.

An exception to the approach above was~\cite{lai10}, 
where chimeras in a network of two populations of Stuart-Landau
oscillators were studied using a self-consistency argument. The existence of a chimera
state was determined from the periodic solution of an ordinary differential equation (ODE),
but this approach did not provide information on the stability or otherwise of the solution
found.

In this paper we use techniques from~\cite{clupol18} to revisit the system studied
in~\cite{lai10} and calculate stability information for the solutions found there.
Since the approach in~\cite{clupol18} is generally applicable to a situation in which
oscillators in one population lie on a closed smooth curve, we then apply these ideas
to three more networks formed from two coupled populations. The second network we
consider consists of Kuramoto oscillators with inertia, each described by a second order
ODE. The third network consists of FitzHugh-Nagumo neural oscillators, each described
by a pair of ODEs. Unlike the oscillators
studied in the first and second networks, these are not invariant under a global
phase shift. The last network we consider consists of Stuart-Landau oscillators
with delayed coupling both within and between populations.

We now briefly present the results from~\cite{clupol18} which we will use.
Sec.~\ref{sec:res} contains the analysis and results for the four types of networks,
while Sec.~\ref{sec:disc} contains a discussion and conclusion.

Clusella and Politi~\cite{clupol18} consider a network of $N$ oscillators,
with the state of the $j$th oscillator being described by the complex variable $z_j$.
The dynamics is given by
\be
   \frac{dz_j}{dt}=f(z_j,\bar{z};K) \label{eq:dzdtorig}
\ee
for some function $f$
where the mean field is given by
\be 
   \bar{z}=\frac{1}{N}\sum_{k=1}^N z_k
\ee
and $K$ is the strength of coupling between an oscillator and the mean field.
For some values of $K$ it is observed that 
when the states of all oscillators are plotted as points in the complex plane, they 
lie on a smooth curve, \mc, enclosing the origin, the shape of which
is parametrised by an angle $\phi$. The distance from the origin to \mc \ at angle
$\phi$ is $R(\phi,t)$, and the density at the point parametrised by
$\phi$ is $P(\phi,t)$. Writing $z_j=r_je^{i\phi_j}$ we can write~\eqref{eq:dzdtorig} as
\begin{align}
   \frac{dr_j}{dt} & =  F(r_j,\phi_j,\bar{z}) \\
   \frac{d\phi_j}{dt} & = G(r_j,\phi_j,\bar{z})
\end{align}
Clusella and Politi~\cite{clupol18} show that
the dynamics of $R$ and $P$ are given by
\begin{align}
   \frac{\p R}{\p t}(\phi,t) & = F(R,\phi,\bar{z})-G(R,\phi,\bar{z})\frac{\p R}{\p \phi} \\
   \frac{\p P}{\p t}(\phi,t) & = -\frac{\p}{\p \phi}\left[P(\phi,t)G(R,\phi,\bar{z})\right] \label{eq:dPdt}
\end{align}
where
\be
   \bar{z}=\int_0^{2\pi}P(\phi,t)R(\phi,t)e^{i\phi}d\phi
\ee
They used these equations to study the splay state and
self-consistent partial synchrony in a network of
Stuart-Landau oscillators. Of course, such equations are only a valid description
of the dynamics of the network if the oscillators do lie on a curve \mc, which should be checked
by solving the original equations governing their dynamics. Numerically, we will treat
$R$ and $P$ as continuous functions of $\phi$, corresponding to an infinite number of oscillators.


While~\cite{clupol18} considered a single population of all-to-all coupled oscillators,
the approach is also valid for a network of two populations of oscillators in which 
oscillators in one population lie on a smooth closed curve while those in the other
population are perfectly synchronous, i.e.~a chimera state.
(It is also valid when oscillators from each population lie on their own curve; 
see Sec.~\ref{sec:FHN}.)

\section{Results}
\label{sec:res}

\subsection{Stuart-Landau oscillators}
\label{sec:SL}
We first consider the chimera state found in~\cite{lai10}. The
equations governing the dynamics are
\begin{align}
   \frac{dX_j}{dt} & =i\omega X_j+\epsilon^{-1}\{1-(1+\delta\epsilon i)|X_j|^2\}X_j+e^{-i\alpha}\left(\frac{\mu}{N}\sum_{k=1}^N X_k+\frac{\nu}{N}\sum_{k=1}^N X_{N+k}\right)
\label{eq:dXdt1}
\end{align}
for $j=1,\dots N$ and
\begin{align}
   \frac{dX_j}{dt} & =i\omega X_j+\epsilon^{-1}\{1-(1+\delta\epsilon i)|X_j|^2\}X_j +e^{-i\alpha}\left(\frac{\mu}{N}\sum_{k=1}^N X_{N+k}+\frac{\nu}{N}\sum_{k=1}^N X_{k}\right)
\label{eq:dXdt2}
\end{align}
for $j=N+1,\dots 2N$, where each $X_j\in\mathbb{C}$ and $\omega,\epsilon,\delta,\alpha,\mu$ and
$\nu$ are all real parameters. 
In a chimera state, $X_j=Y$ for $j\in\{N+1,\dots 2N\}$, i.e.~population two is perfectly
synchronised. Letting
\be
   \bar{X}=\frac{1}{N}\sum_{k=1}^N X_k
\ee
we have
\be
   \frac{dY}{dt}  =i\omega Y+\epsilon^{-1}\{1-(1+\delta\epsilon i)|Y|^2\}Y
+e^{-i\alpha}\left(\mu Y+\nu\bar{X}\right)
\ee
and each oscillator in population one satisfies
\begin{align}
   \frac{dX_j}{dt} & =i\omega X_j+\epsilon^{-1}\{1-(1+\delta\epsilon i)|X_j|^2\}X_j +e^{-i\alpha}\left(\mu \bar{X}+\nu Y\right),
\end{align}
for $j=1,\dots N$. Writing $X_j=r_je^{i\phi_j}$ we have
\begin{align}
   \frac{dr_j}{dt} & = \epsilon^{-1}(1-r_j^2)r_j+\mbox{Re}\left[e^{-i(\alpha+\phi_j)}\left(\mu \bar{X}+\nu Y\right)\right] \equiv F(r_j,\phi_j,\bar{X},Y) \\
   \frac{d\phi_j}{dt} & = \omega-\delta r_j^2+\frac{1}{r_j}\mbox{Im}\left[e^{-i(\alpha+\phi_j)}\left(\mu \bar{X}+\nu Y\right)\right]\equiv G(r_j,\phi_j,\bar{X},Y) \label{eq:G}
\end{align}
Thus we consider the dynamical system
\begin{align}
   \frac{\p R}{\p t}(\phi,t) & = F(R,\phi,\bar{X},Y)-G(R,\phi,\bar{X},Y)\frac{\p R}{\p \phi} \label{eq:dRdt} \\
   \frac{\p P}{\p t}(\phi,t) & = -\frac{\p}{\p \phi}\left[P(\phi,t)G(R,\phi,\bar{X},Y)\right]+D\frac{\p^2}{\p \phi^2}P(\phi,t) \\
    \frac{dY}{dt} & =i\omega Y+\epsilon^{-1}\{1-(1+\delta\epsilon i)|Y|^2\}Y+e^{-i\alpha}\left(\mu Y+\nu\bar{X}\right) \label{eq:dYdt}
\end{align}
where
\be
   \bar{X}=\int_0^{2\pi}P(\phi,t)R(\phi,t)e^{i\phi}d\phi \label{eq:Xb}
\ee
and for numerical stability reasons we have added a small amount of diffusion, of strength $D$,
to~\eqref{eq:dPdt} (as did~\cite{clupol18}). The equations~\eqref{eq:dRdt}-\eqref{eq:Xb}
form a coupled PDE/ODE system.
We define $\beta=\pi/2-\alpha$ and let $\mu=(1+A)/2,\nu=(1-A)/2$.

Note that~\eqref{eq:dXdt1}-\eqref{eq:dXdt2} are invariant under the global phase shift
$X_j\mapsto X_je^{i\gamma}$ for any constant $\gamma$ and thus we can move to a rotating
coordinate frame in which $Y$ is constant, and we can then shift our coordinate system so that
$Y$ is real. Moving to a coordinate frame rotating with speed $\Omega$
has the effect of replacing $\omega$ in~\eqref{eq:G} and~\eqref{eq:dYdt}  by
$\omega+\Omega$.

\begin{figure}
\begin{center}
\includegraphics[width=13cm]{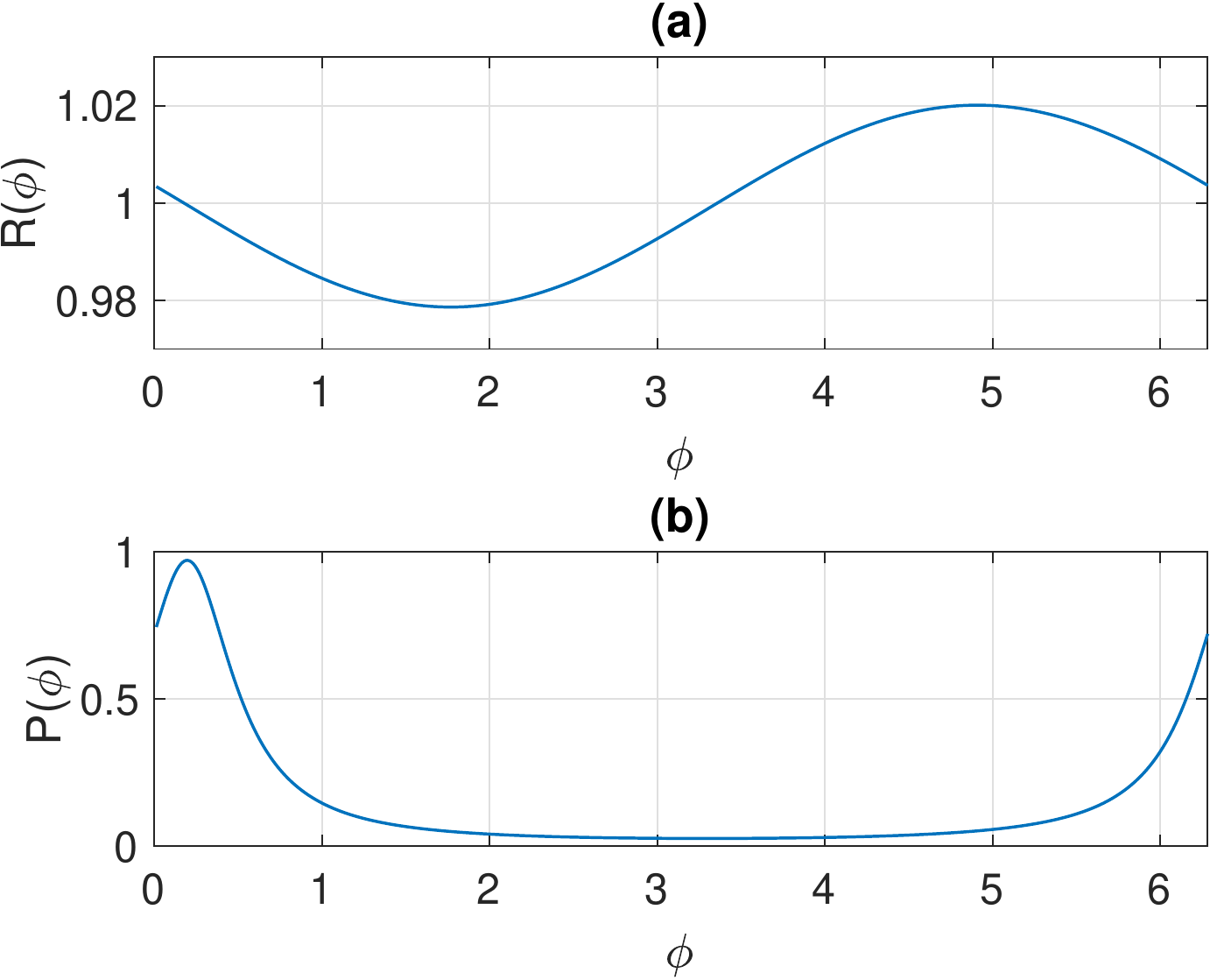}
\caption{Snapshot of a solution of~\eqref{eq:dRdt}-\eqref{eq:Xb} for which $Y$ is real.
(a): $R(\phi)$, (b): $P(\phi)$.
Parameters: $\epsilon=0.05,\omega=0,\delta=-0.01,A=0.2,\beta=0.08,D=10^{-8}$.}
\label{fig:snapP}
\end{center}
\end{figure}

\begin{figure}
\begin{center}
\includegraphics[width=13cm]{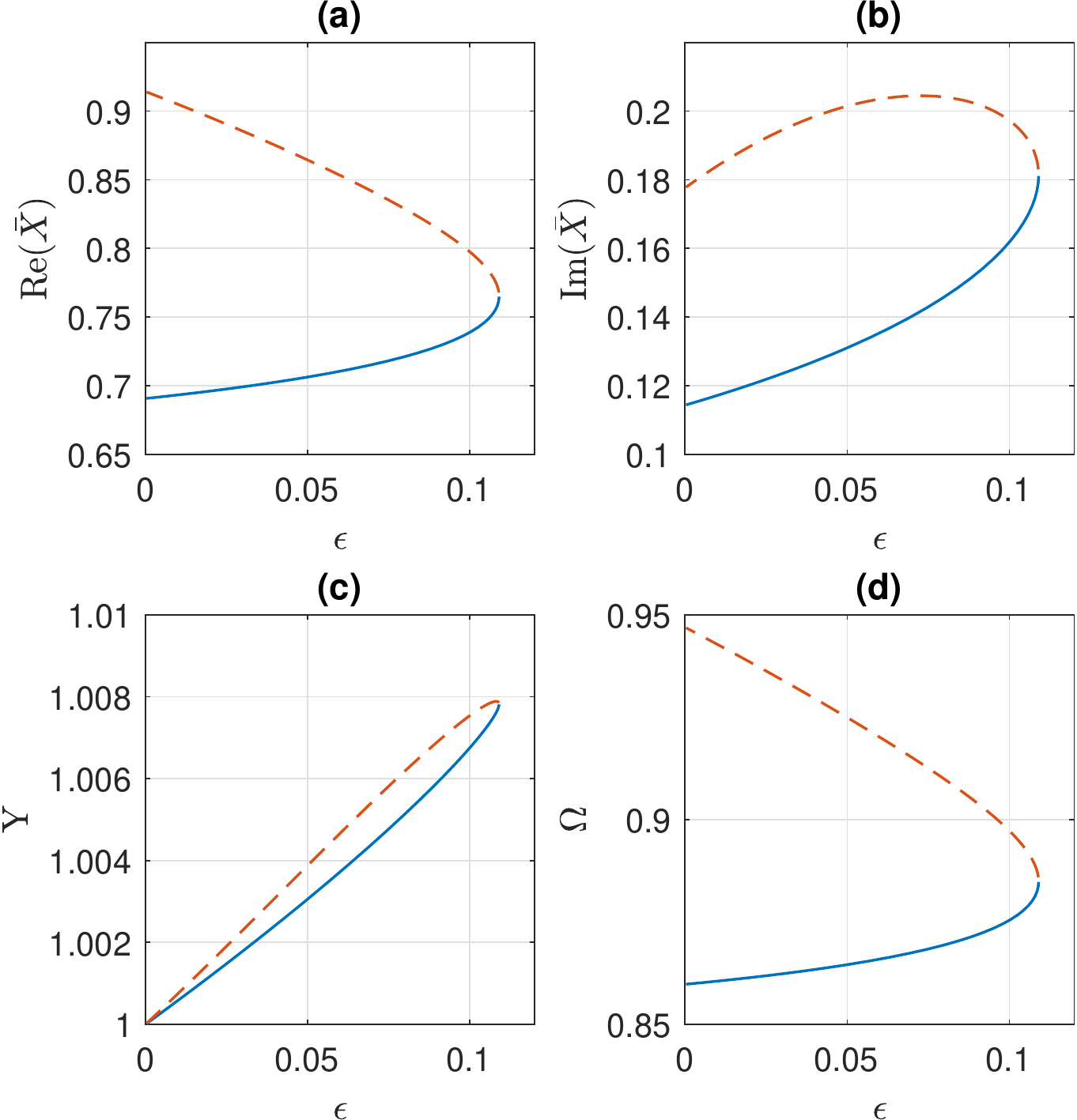}
\caption{Steady states of~\eqref{eq:dRdt}-\eqref{eq:Xb} as functions
of $\epsilon$.
(a): Re($\bar{X}$), (b): Im($\bar{X}$), (c): $Y$ and (d): $\Omega$.
Solid: stable; dashed: unstable.
Other parameters: $\omega=0,\delta=-0.01,A=0.2,\beta=0.08,D=10^{-8}$.}
\label{fig:varyep}
\end{center}
\end{figure}

We numerically integrate~\eqref{eq:dRdt}-\eqref{eq:Xb} in time to find a stable solution.
An example is shown in Fig.~\ref{fig:snapP}. (Compare with Fig.~1 of~\cite{lai10}.)
We discretised $\phi$ using
$256$ equally-spaced points and implemented derivatives with respect to $\phi$
spectrally~\cite{tre00}. We enforce conservation of probability by setting $P$ at one
grid point equal to $1/\Delta$ minus the sum of the values at all other grid points,
where $\Delta=2\pi/256$, the $\phi$ grid spacing~\cite{erm06}. 
We then
follow the solution in Fig.~\ref{fig:snapP} using pseudo-arclength continuation~\cite{lai14,auto}
as $\epsilon$ is varied. 
The results are shown in Fig.~\ref{fig:varyep}, and we have reproduced the first four panels in Fig.~2 of~\cite{lai10}. Note that stability is calculated from the eigenvalues of the linearisation
of~\eqref{eq:dRdt}-\eqref{eq:Xb} about the steady state, unlike in~\cite{lai10} where it was
just inferred.

\begin{figure}
\begin{center}
\includegraphics[width=13cm]{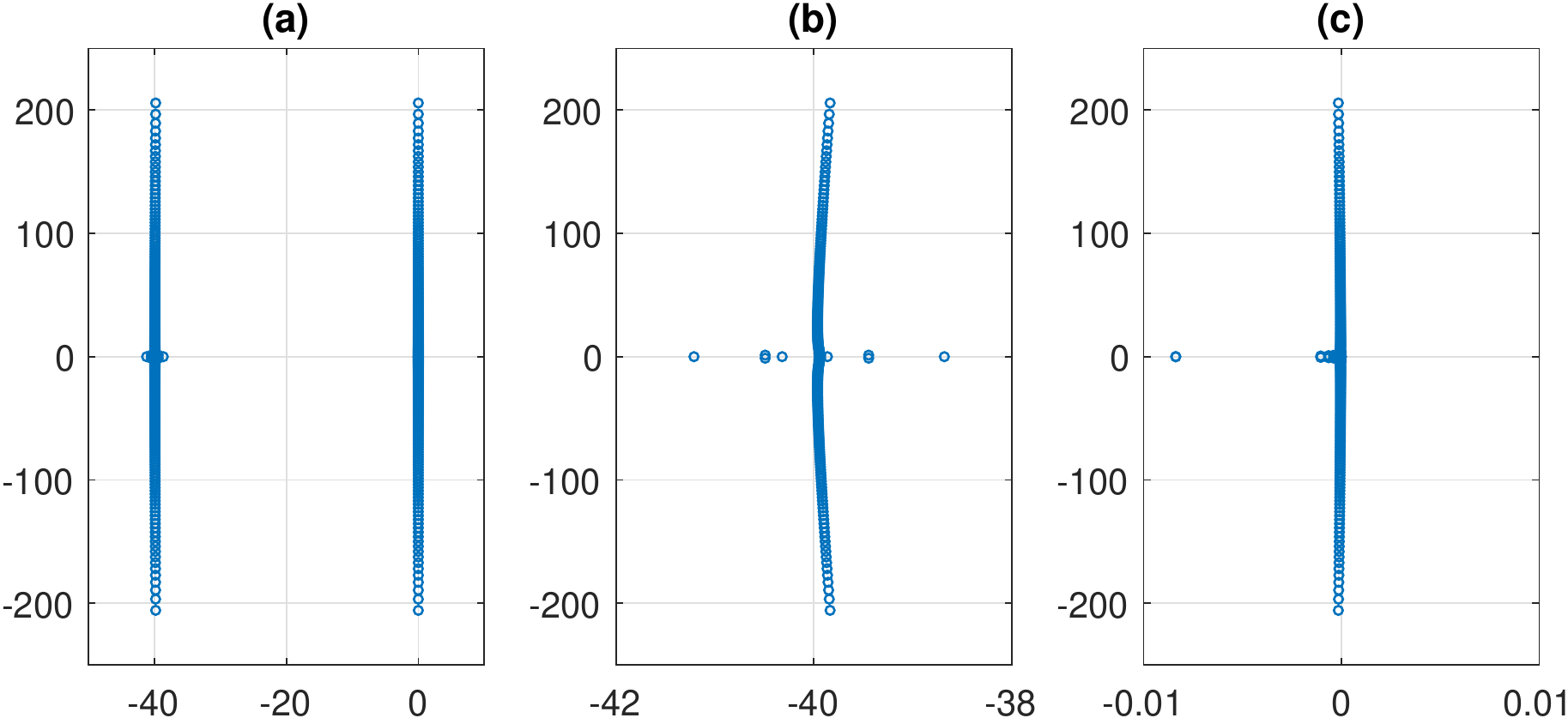}
\caption{(a): spectrum of the solution shown in Fig.~\ref{fig:snapP}. Panels (b) and
(c) show details of the two clusters of eigenvalues.
Parameters: $\epsilon=0.05,\omega=0,\delta=-0.01,A=0.2,\beta=0.08,D=10^{-8}$.}
\label{fig:spec}
\end{center}
\end{figure}

The eigenvalues, $\lambda_j$, of the linearisation of~\eqref{eq:dRdt}-\eqref{eq:Xb} about the
solution shown in Fig.~\ref{fig:snapP} are plotted in the complex plane in Fig.~\ref{fig:spec}.
We notice that they form two clusters, one around Re$(\lambda_j)=-40$ and the other
around  Re$(\lambda_j)=0$. The first group can be understood by linearising
$F$ with respect to $R$. We obtain $\epsilon^{-1}(1-3R^2)$, and evaluating this at $R=1$ gives
$-2/\epsilon=-40$, for this solution. The second group of eigenvalues is presumably
related to the dynamics of $P$, and has been observed in other 
similar systems~\cite{clupol16,clupol18,ome18}. The slight
deviation from the imaginary axis visible in panel (c) of Fig.~\ref{fig:spec} is due to the
non-zero value of $D$ used ($D=10^{-8}$). If $D$ is set to zero when calculating the
eigenvalues, this group lies very close to the imaginary axis ($|\mbox{Re}(\lambda_j)|<10^{-9}$).

As mentioned in~\cite{clupol18}, one could find a steady state of~\eqref{eq:dRdt}-\eqref{eq:Xb}
with $D=0$ 
by assuming a value for $\bar{X}$, solving $dY/dt=0$ for $Y$, numerically integrating
\be
   \frac{\partial R}{\partial \phi}=\frac{F(R,\phi,\bar{X},Y)}{G(R,\phi,\bar{X},Y)}
\ee
to obtain $R_0(\phi)$, setting
\be
  P_0(\phi)=\frac{\eta}{G(R_0,\phi,\bar{X},Y)}
\ee
where $\eta$ is a normalisation constant (since $P$ is a probability density) and then
requiring that
\be
   \int_0^{2\pi}P_0(\phi)R_0(\phi)e^{i\phi}d\phi 
\ee
is equal to the value originally assumed for $\bar{X}$. Such an approach is equivalent to
that taken in~\cite{lai10}, where the equations governing a single
oscillator in population one 
\begin{align}
   \frac{dr}{dt} & = F(r,\phi,\bar{X},Y) \label{eq:drdt} \\
   \frac{d\phi}{dt} & = G(r,\phi,\bar{X},Y) \label{eq:dphidt}
\end{align}
were numerically solved in a self-consistent way to show the existence of a chimera.

Now each oscillator in population one satisfies~\eqref{eq:drdt}-\eqref{eq:dphidt}.
Thus having found a steady state of~\eqref{eq:dRdt}-\eqref{eq:Xb} by integrating
these equations in time, we can find a periodic
solution of~\eqref{eq:drdt}-\eqref{eq:dphidt}. $2\pi$ divided by the period of this
orbit then gives the angular frequency of an incoherent oscillator, relative to that of the
synchronous group (whose frequency in the original coordinate frame is $\Omega$).
For all of the points shown in Fig.~\ref{fig:varyep},~\eqref{eq:drdt}-\eqref{eq:dphidt}
has a {\em stable} periodic solution, the period of which is shown in Fig.~\ref{fig:TSL}.
Note that this Figure reproduces panel (e) in Fig.~2 of~\cite{lai10}.

\begin{figure}
\begin{center}
\includegraphics[width=13cm]{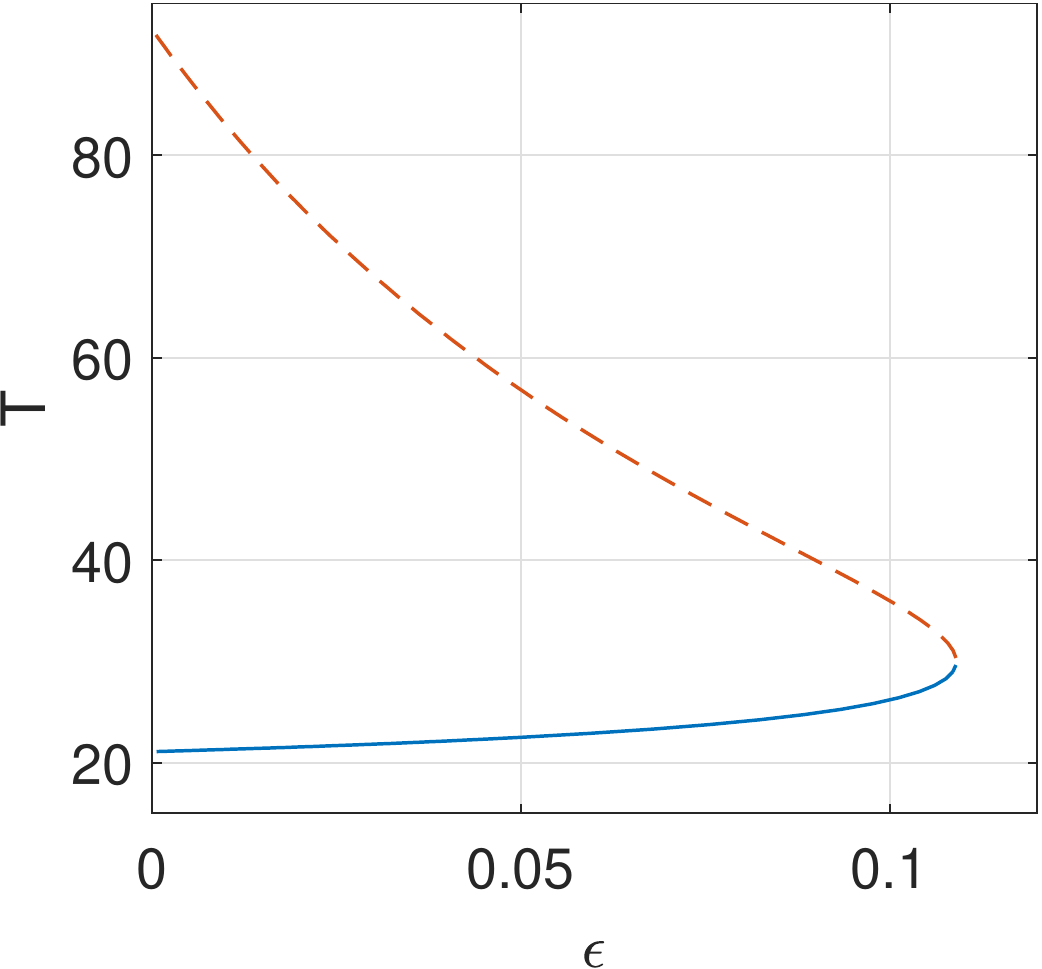}
\caption{Period, $T$, of the stable periodic solution of~\eqref{eq:drdt}-\eqref{eq:dphidt},
where the values of $\bar{X},Y$ and $\Omega$ are those shown in Fig.~\ref{fig:varyep}.
Solid and dashed lines refer to stability indicated in Fig.~\ref{fig:varyep}.
Other parameters: $\omega=0,\delta=-0.01,A=0.2,\beta=0.08,D=10^{-8}$.}
\label{fig:TSL}
\end{center}
\end{figure}

Following the saddle-node bifurcation shown in Fig.~\ref{fig:varyep} as $A$ is varied we
obtain Fig.~\ref{fig:TB}. By increasing $A$ for $\epsilon=0.05$ we find a Hopf bifurcation,
also shown in Fig.~\ref{fig:TB}. Numerical investigations suggest that this bifurcation
is supercritical, and that the oscillations created in it are destroyed in a homoclinic
bifurcation to the right of the Hopf curve in Fig.~\ref{fig:TB}. The curve of homoclinic
bifurcations should terminate at the codimension-two point where the Hopf curve meets
the saddle-node curve; this
scenario is observed in many systems showing 
chimeras~\cite{lai10,lai09,abrmir08,panabr16,mar10,mar10a,marbic16}.
Note that the curve of Hopf bifurcations was found by following
the algebraic equations defining such a bifurcation, whereas in~\cite{lai10}, such a curve was found
through direct simulation of~\eqref{eq:dXdt1}-\eqref{eq:dXdt2}.

\begin{figure}
\begin{center}
\includegraphics[width=13cm]{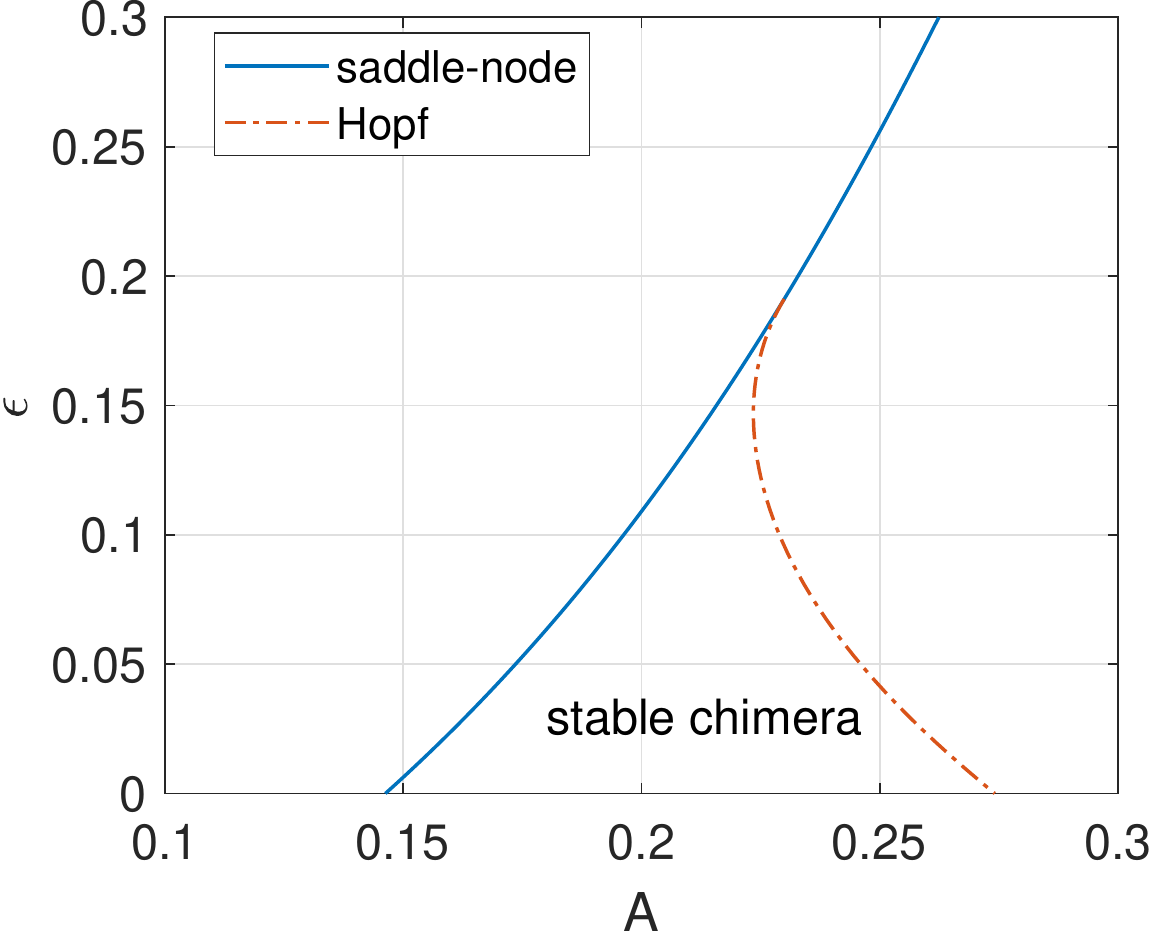}
\caption{Continuation of the saddle-node bifurcation shown in Fig.~\ref{fig:varyep} (solid)
and a Hopf bifurcation (dash-dotted). Oscillating chimeras exist slightly to the right
of the Hopf curve.
Other parameters: $\omega=0,\delta=-0.01,\beta=0.08,D=10^{-8}$.}
\label{fig:TB}
\end{center}
\end{figure}  

We end this section by noting that with the approach presented here 
we cannot detect bifurcations in which the synchronous group becomes asynchronous.
Also, above the saddle-node curve in Fig.~\ref{fig:TB} the only attractor is the fully
synchronous state and the approach presented here cannot be used to study this state, as $P$
approaches a delta function in $\phi$ and our numerical scheme breaks down.

\subsection{Kuramoto with inertia}
\label{sec:inertia}
\begin{figure}
\begin{center}
\includegraphics[width=13cm]{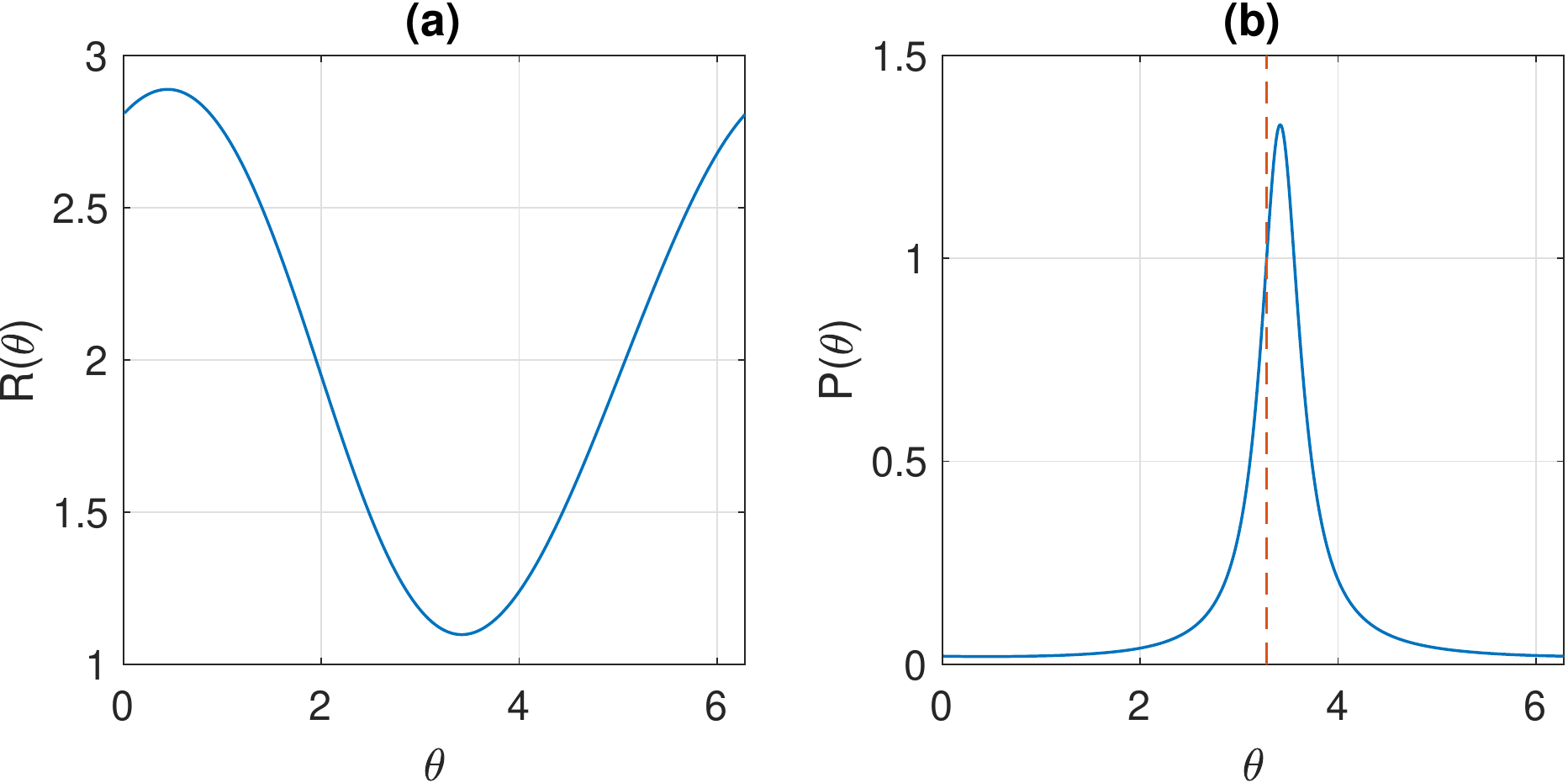}
\caption{Steady state of~\eqref{eq:dRpend}-\eqref{eq:dPpend},~\eqref{eq:dUdtA} 
and~\eqref{eq:dThetadtA}.
(a): $R(\theta)$. (b): $P(\theta)$ (solid) with the value of $\Theta$ shown dotted.
Parameters: $\mu=0.6,\nu=0.4,\alpha=\pi/2-0.05,m=0.1,D=10^{-4}$.}
\label{fig:snappend}
\end{center}
\end{figure}

We now consider a network formed from two populations of $N$ Kuramoto oscillators with inertia. 
The system is described by
\begin{align}
   m\frac{d^2\theta_i^{(1)}}{dt^2}+\frac{d\theta_i^{(1)}}{dt} & =\omega+\frac{\mu}{N}\sum_{j=1}^N\sin{\left(\theta_j^{(1)}-\theta_i^{(1)}-\alpha\right)} +\frac{\nu}{N}\sum_{j=1}^N\sin{\left(\theta_j^{(2)}-\theta_i^{(1)}-\alpha\right)} \label{eq:pendA} \\
   m\frac{d^2\theta_i^{(2)}}{dt^2}+\frac{d\theta_i^{(2)}}{dt} & =\omega+\frac{\mu}{N}\sum_{j=1}^N\sin{\left(\theta_j^{(2)}-\theta_i^{(2)}-\alpha\right)} +\frac{\nu}{N}\sum_{j=1}^N\sin{\left(\theta_j^{(1)}-\theta_i^{(2)}-\alpha\right)} \label{eq:pendB}
\end{align}
where $m$ is ``mass'', $\omega,\mu,\nu$ and $\alpha$ are parameters,
and the superscript labels the population. When $m=0$
this reverts to a previously studied
case~\cite{abrmir08,pikros08}. It is reasonable to expect that chimeras may exist and be
stable for $m$ in some interval $[0,m_0]$, as found via numerical simulations of
slightly heterogeneous oscillators~\cite{boukan14}.
Note that the system is invariant under a uniform shift
of all of the phases, so we can set $\omega=0$ without loss of generality.
We rewrite the equations as
\begin{align}
   \frac{d\theta_i^{(1)}}{dt} & = u_i^{(1)} \\
   \frac{du_i^{(1)}}{dt} & = \left[-u_i^{(1)}+\frac{\mu}{N}\sum_{j=1}^N\sin{\left(\theta_j^{(1)}-\theta_i^{(1)}-\alpha\right)} +\frac{\nu}{N}\sum_{j=1}^N\sin{\left(\theta_j^{(2)}-\theta_i^{(1)}-\alpha\right)}\right]/m \\
    \frac{d\theta_i^{(2)}}{dt} & = u_i^{(2)} \\
   \frac{du_i^{(2)}}{dt} & = \left[-u_i^{(2)}+\frac{\mu}{N}\sum_{j=1}^N\sin{\left(\theta_j^{(2)}-\theta_i^{(2)}-\alpha\right)} +\frac{\nu}{N}\sum_{j=1}^N\sin{\left(\theta_j^1-\theta_i^{(2)}-\alpha\right)}\right]/m
\end{align}
In a chimera state let us assume that
population two is fully synchronised, with $\theta_i^{(2)}=\Theta$ for $i=1,2\dots N$.
This population satisfies
\begin{align}
   \frac{d\Theta}{dt} & = U  \label{eq:dThetadt} \\
   \frac{dU}{dt} & = \left[-U-\mu\sin{\alpha}+\frac{\nu}{N}\sum_{j=1}^N\sin{(\theta_j-\Theta-\alpha)}\right]/m \nonumber \\
   & = \left[-U-\mu\sin{\alpha}+\nu\mbox{Im}\left\{e^{-i(\Theta+\alpha)}X\right\}\right]/m \label{eq:dUdtA}
\end{align}
where
\be
   X\equiv\frac{1}{N}\sum_{j=1}^N e^{i\theta_j}\in\mathbb{C},
\ee
the sums are over population one,
and we have dropped the superscripts. Oscillators in population one satisfy
\begin{align}
   \frac{d\theta_i}{dt} & = u_i \\
   \frac{du_i}{dt} & = \left[-u_i+\frac{\mu}{N}\sum_{j=1}^N\sin{(\theta_j-\theta_i-\alpha)}+\nu\sin{(\Theta-\theta_i-\alpha)}\right]/m \nonumber \\
    & = \left[-u_i+\mu\mbox{Im}\left\{e^{-i(\theta_i+\alpha)}X\right\}+\nu\sin{(\Theta-\theta_i-\alpha)}\right]/m
\end{align}
We put these equations in ``polar'' form by defining $r_j=2+u_j$ and thus we have
\begin{align}
   \frac{dr_j}{dt} & = \left[-(r_j-2)+\mu\mbox{Im}\left\{e^{-i(\theta_i+\alpha)}X\right\} +\nu\sin{(\Theta-\theta_i-\alpha)}\right]/m\equiv F(r_j,\theta_j,X,\Theta) \\
   \frac{d\theta_j}{dt} & = r_j-2 \label{eq:dthetadt}
\end{align} 
The chimera state of interest is stationary in a coordinate frame rotating at speed $\Omega$.
Moving to this coordinate frame has the effect of replacing~\eqref{eq:dThetadt} by 
\be
   \frac{d\Theta}{dt} = U+\Omega \label{eq:dThetadtA}
\ee
and~\eqref{eq:dthetadt} by
\be
   \frac{d\theta_j}{dt}= r_j-2+\Omega \equiv G(r_j,\theta_j,X,\Theta)
\ee
Thus we consider the dynamical system
\begin{align}
   \frac{\p R}{\p t}(\theta,t) & = F(R,\theta,X,\Theta)-G(R,\theta,X,\Theta)\frac{\p R}{\p \theta} \label{eq:dRpend} \\
   \frac{\p P}{\p t}(\theta,t) & = -\frac{\p}{\p \theta}\left[P(\theta,t)G(R,\theta,X,\Theta)\right]+D\frac{\p^2}{\p \theta^2}P(\theta,t) \label{eq:dPpend}
\end{align}
along with~\eqref{eq:dUdtA} and~\eqref{eq:dThetadtA}
where
\be
   X(t)=\int_0^{2\pi}P(\theta,t)R(\theta,t)e^{i\theta}d\theta \label{eq:X}
\ee
Choosing parameters $\mu=0.6,\nu=0.4,\alpha=\pi/2-0.05,m=0.1,D=10^{-4}$ and numerically
integrating this system
we find a stable steady state, shown in Fig.~\ref{fig:snappend}. However, decreasing $D$ from
this value we find that this solution is actually unstable for smaller values of $D$, with the
instability seeming to be a Hopf bifurcation.

\begin{figure}
\begin{center}
\includegraphics[width=13cm]{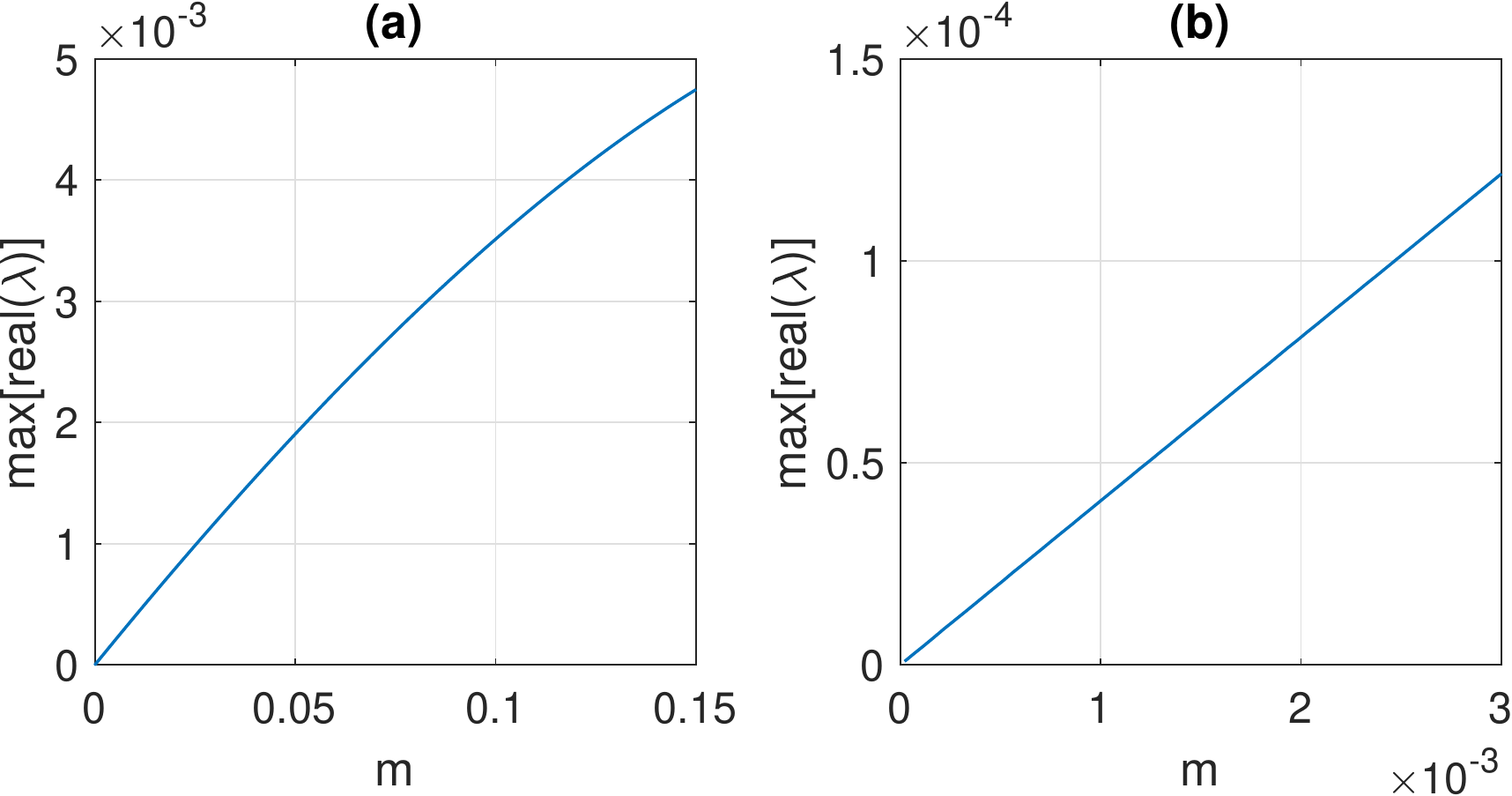}
\caption{(a): Maximum of the real parts of the eigenvalues of the linearisation about a 
steady state of~\eqref{eq:dRpend}-\eqref{eq:dPpend},~\eqref{eq:dUdtA} 
and~\eqref{eq:dThetadtA}, as a function of $m$. (b): Zoom of panel (a).
Other parameters: $\mu=0.6,\nu=0.4,\alpha=\pi/2-0.05,D=10^{-13}$.}
\label{fig:eigpend}
\end{center}
\end{figure}

To verify this we followed the steady state shown in Fig.~\ref{fig:snappend} as $m$ was varied,
for a very small value of $D$ ($D=10^{-13}$). The real part of the right-most eigenvalues
of the linearisation about this state are shown in Fig.~\ref{fig:eigpend}, and we see that
for all values of $m$, these are positive (and the right-most eigenvalues are a complex
conjugate pair). Thus the system~\eqref{eq:dRpend}-\eqref{eq:dPpend},~\eqref{eq:dUdtA} 
and~\eqref{eq:dThetadtA} does not support a stable chimera for small values of $m$.
(The system was discretised in $\theta$ using $256$ equally-spaced grid points. Doubling
this number did not change the results obtained. 
The paper~\cite{boukan14} does, however, show numerical evidence of the existence of
stable chimeras in a {\em finite} network of {\it heterogeneous} oscillators of the
form~\eqref{eq:pendA}-\eqref{eq:pendB} for small $m$ and the same values of other 
parameters as used here.

Olmi~\cite{olm15} considered~\eqref{eq:pendA}-\eqref{eq:pendB} for $\omega=1,N=200,\alpha=\pi/2-0.02$,
and the same values of $\mu$ and $\nu$ as used here. Repeating the analysis above for this
value of $\alpha$ we find qualitatively the same picture as that shown in Fig.~\ref{fig:eigpend}.
Olmi observed that even for $m=10^{-4}$, oscillations in 
the magnitude of the order parameter of the asynchronous
population grew, ``but over very long times scales,'' consistent with our results.
Repeating the calculations shown in
Fig.~\ref{fig:eigpend} but for $\alpha=\pi/2-0.02$, then interpolating to find the real part of
the rightmost eigenvalues for $m=10^{-4}$, we obtain $\sim 4.8\times 10^{-6}$. Thus over 
$5\times 10^5$ time units, we expect the amplitude of these fluctuations to grow by a factor of 
approximately $\exp{(4.8\times 10^{-6} \times 5\times 10^5)}\approx 11$, in excellent agreement
with Olmi's observation of growth by a factor of $10$.

In conclusion, stable chimera states (stationary in a uniformly rotating frame)
do not exist in~\eqref{eq:pendA}-\eqref{eq:pendB}
for infinite $N$ for {\em any} small values of $m$ using the values of other parameters 
from~\cite{boukan14}, or from~\cite{olm15}.

\subsection{FitzHugh-Nagumo oscillators}
\label{sec:FHN}

In this section we consider two populations of FitzHugh-Nagumo oscillators.
In~\cite{omeome13} the authors considered a ring of such oscillators, nonlocally coupled,
and showed numerically that such a system could support chimeras.

\begin{figure}
\begin{center}
\includegraphics[width=13cm]{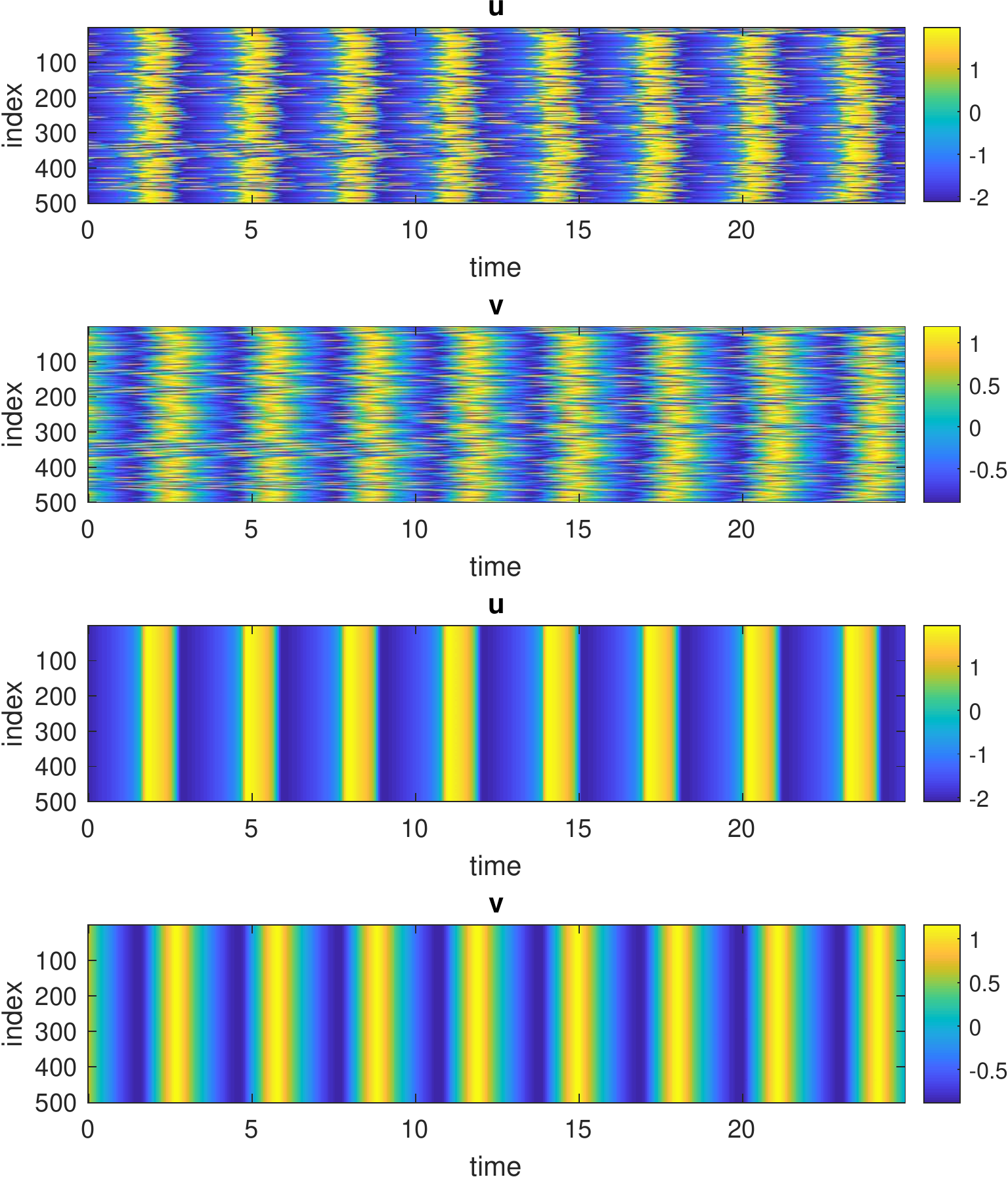}
\caption{A chimera state for equations~\eqref{eq:dudtFHN}-\eqref{eq:dvdtFHN}.
The top two rows show $u$ and $v$ for population one, and the bottom two 
show them for population two.
Parameters: $N=500,\phi=\pi/2-0.1,\epsilon=0.1,a=0.5,A=0.2$.}
\label{fig:FHN}
\end{center}
\end{figure}

Consider the following network:
\begin{align}
   \epsilon\frac{du_i}{dt} & = u_i-u_i^3/3-v_i+\mu\left[b_{uu}\left(U_1-u_i\right)+b_{uv}\left(V_1-v_i\right)\right] \nonumber \\
 & +\nu\left[b_{uu}\left(U_2-u_i\right)+b_{uv}\left(V_2-v_i\right)\right] \label{eq:dudtFHN} \\
   \frac{dv_i}{dt} & = u_i+a+\mu\left[b_{vu}\left(U_1-u_i\right)+b_{vv}\left(V_1-v_i\right)\right] +\nu\left[b_{vu}\left(U_2-u_i\right)+b_{vv}\left(V_2-v_i\right)\right]
\end{align}
for $i=1,2\dots N$ and
\begin{align}
   \epsilon\frac{du_i}{dt} & = u_i-u_i^3/3-v_i+\mu\left[b_{uu}\left(U_2-u_i\right)+b_{uv}\left(V_2-v_i\right)\right] \nonumber \\
   & +\nu\left[b_{uu}\left(U_1-u_i\right)+b_{uv}\left(V_1-v_i\right)\right] \\
   \frac{dv_i}{dt} & = u_i+a+\mu\left[b_{vu}\left(U_2-u_i\right)+b_{vv}\left(V_2-v_i\right)\right] +\nu\left[b_{vu}\left(U_1-u_i\right)+b_{vv}\left(V_1-v_i\right)\right] \label{eq:dvdtFHN}
\end{align}
for $i=N+1,\dots 2N$, where
\be
   U_1\equiv\frac{1}{N}\sum_{i=1}^N u_i; \qquad V_1\equiv\frac{1}{N}\sum_{i=1}^N v_i
\ee
and
\be
\qquad U_2\equiv\frac{1}{N}\sum_{i=1}^N u_{N+i}; \qquad V_2\equiv\frac{1}{N}\sum_{i=1}^N v_{N+i}
\ee
We have
\be
   \begin{pmatrix} b_{uu} & b_{uv} \\ b_{vu} & b_{vv} \end{pmatrix}
=\begin{pmatrix} \cos{\phi} & \sin{\phi} \\ -\sin{\phi} & \cos{\phi} \end{pmatrix}
\ee
for some phase $\phi$.
Coupling within each population has strength $\mu$ and that between populations
has strength $\nu$, and we control their relative strength by defining $\mu=(1+A)/20$ and
$\nu=(1-A)/20$. Numerically, we find a stable chimera state 
 for $\phi=\pi/2-0.1,\epsilon=0.1,a=0.5,A=0.2$, as shown in Fig.~\ref{fig:FHN}.
Since $\epsilon$ is small the oscillators are relaxation oscillators, with strongly
nonlinear waveforms.

Since each oscillator rotates around the origin, we can define an average angular velocity
by counting the number of rotations each one makes during a long time interval and dividing
by the duration of that interval~\cite{omeome13}. Doing so we find that for these parameter
values the 
asynchronous group has average
angular velocity $\omega=2.1516$ while the synchronous group has $\omega=2.0511$.
While these values are close, the fact that they are different shows that this is a
chimera state. 

\begin{figure}
\begin{center}
\includegraphics[width=13cm]{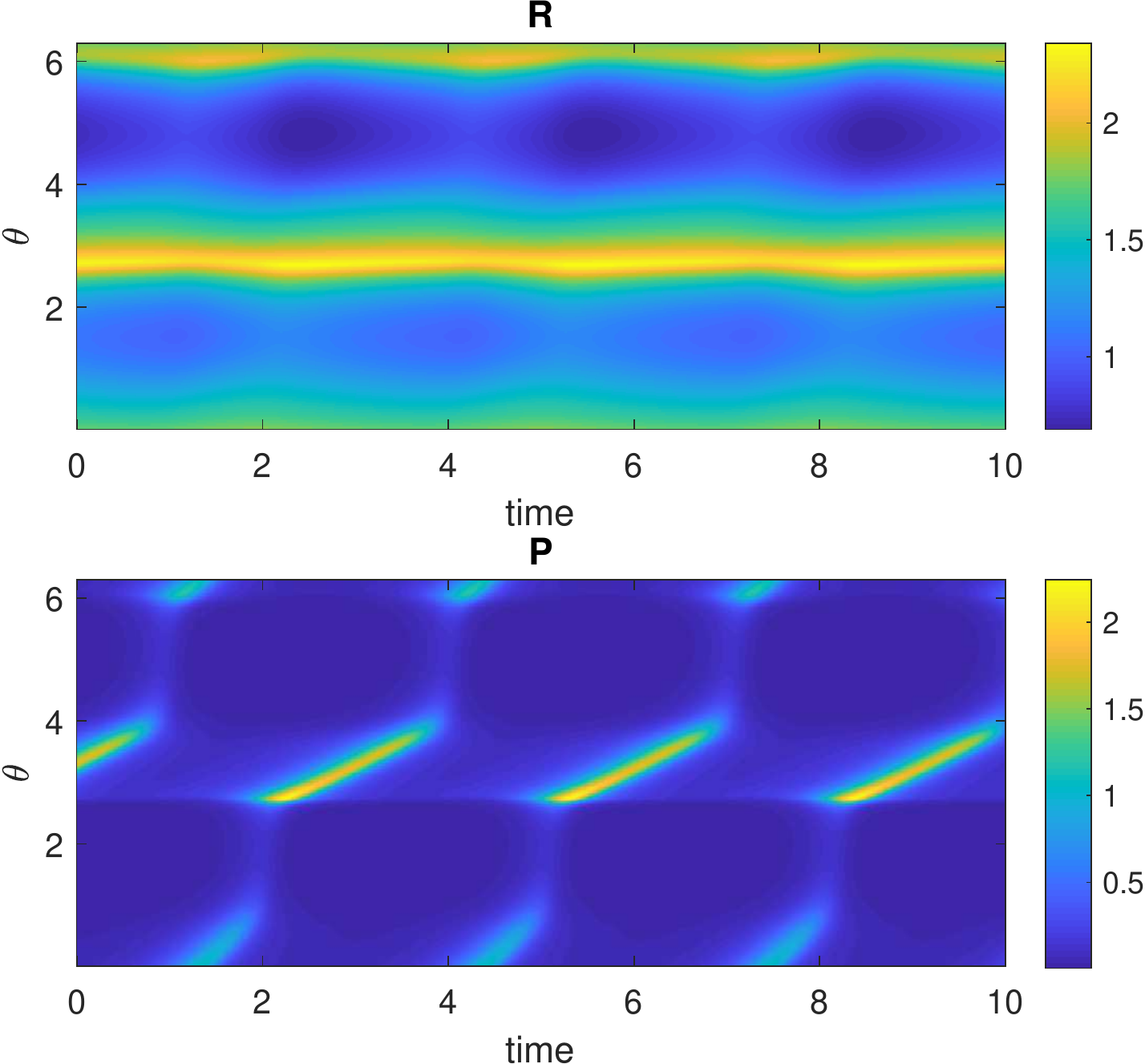}
\caption{A chimera state for equations~\eqref{eq:dRdtFHN}-\eqref{eq:dPdtFHN} and~\eqref{eq:dUdt}-\eqref{eq:dVdt}.
The top row shows $R(\theta,t)$ and the bottom one $P(\theta,t)$.
Parameters: $\phi=\pi/2-0.1,\epsilon=0.1,a=0.5,A=0.2,D=10^{-3}$.
($\theta$ is discretised with 256 points.)}
\label{fig:FHNsol}
\end{center}
\end{figure}

Suppose population two is synchronised. Its dynamics is described by
\begin{align}
   \epsilon\frac{dU_2}{dt} & = U_2-U_2^3/3-V_2+\nu\left[b_{uu}\left(U_1-U_2\right)+b_{uv}\left(V_1-V_2\right)\right] \label{eq:dUdt} \\
   \frac{dV_2}{dt} & = U_2+a+\nu\left[b_{vu}\left(U_1-U_2\right)+b_{vv}\left(V_1-V_2\right)\right] \label{eq:dVdt}
\end{align}
In population one we have
\begin{align}
   \epsilon\frac{du_i}{dt} & = u_i-u_i^3/3-v_i+\mu\left[b_{uu}\left(U_1-u_i\right)+b_{uv}\left(V_1-v_i\right)\right] \nonumber \\
    & +\nu\left[b_{uu}\left(U_2-u_i\right)+b_{uv}\left(V_2-v_i\right)\right] \\
   \frac{dv_i}{dt} & = u_i+a+\mu\left[b_{vu}\left(U_1-u_i\right)+b_{vv}\left(V_1-v_i\right)\right] \nonumber \\
  & +\nu\left[b_{vu}\left(U_2-u_i\right)+b_{vv}\left(V_2-v_i\right)\right]
\end{align}
for $i=1,2\dots N$. Writing $r_i^2=u_i^2+v_i^2$ and $\tan{\theta_i}=v_i/u_i$ so that
$u_i=r_i\cos{\theta_i}$ and $v_i=r_i\sin{\theta_i}$ we have
\begin{align}
	   \frac{dr_i}{dt} & =\frac{u_i\frac{du_i}{dt}+v_i\frac{dv_i}{dt}}{r_i}\equiv F(r_i,\theta_i,U_1,V_1,U_2,V_2) \\
    \frac{d\theta_i}{dt} & = \frac{u_i\frac{dv_i}{dt}-v_i\frac{du_i}{dt}}{r_i^2}\equiv G(r_i,\theta_i,U_1,V_1,U_2,V_2)
\end{align}
Thus we consider the dynamical system
\begin{align}
   \frac{\p R}{\p t}(\theta,t) & = F(R,\theta,U_1,V_1,U_2,V_2) -G(R,\theta,U_1,V_1,U_2,V_2)\frac{\p R}{\p \theta}+D\frac{\p^2}{\p \theta^2}R(\theta,t) \label{eq:dRdtFHN} \\
   \frac{\p P}{\p t}(\theta,t) & = -\frac{\p}{\p \theta}\left[P(\theta,t)G(R,\theta,U_1,V_1,U_2,V_2)\right]+D\frac{\p^2}{\p \theta^2}P(\theta,t) \label{eq:dPdtFHN}
\end{align}
together with~\eqref{eq:dUdt}-\eqref{eq:dVdt},
where
\be
   U_1=\int_0^{2\pi}P(\theta,t)R(\theta,t)\cos{\theta}d\theta \label{eq:PQa}
\ee
and
\be
  V_1=\int_0^{2\pi}P(\theta,t)R(\theta,t)\sin{\theta}d\theta \label{eq:PQb}
\ee
and we have added a small amount of
diffusion in both~\eqref{eq:dRdtFHN}-\eqref{eq:dPdtFHN} to stabilise
solutions. A significant difference between the system studied in this section and those
in Secs.~\ref{sec:SL} and~\ref{sec:inertia} (and~\ref{sec:del}, below)
is that the FitzHugh-Nagumo system is not invariant
under a global phase shift. Thus the chimera state of interest is a stable periodic solution
of~\eqref{eq:dRdtFHN}-\eqref{eq:dPdtFHN} and~\eqref{eq:dUdt}-\eqref{eq:dVdt}, as seen in
Fig.~\ref{fig:FHNsol}.

\begin{figure}
\begin{center}
\includegraphics[width=13cm]{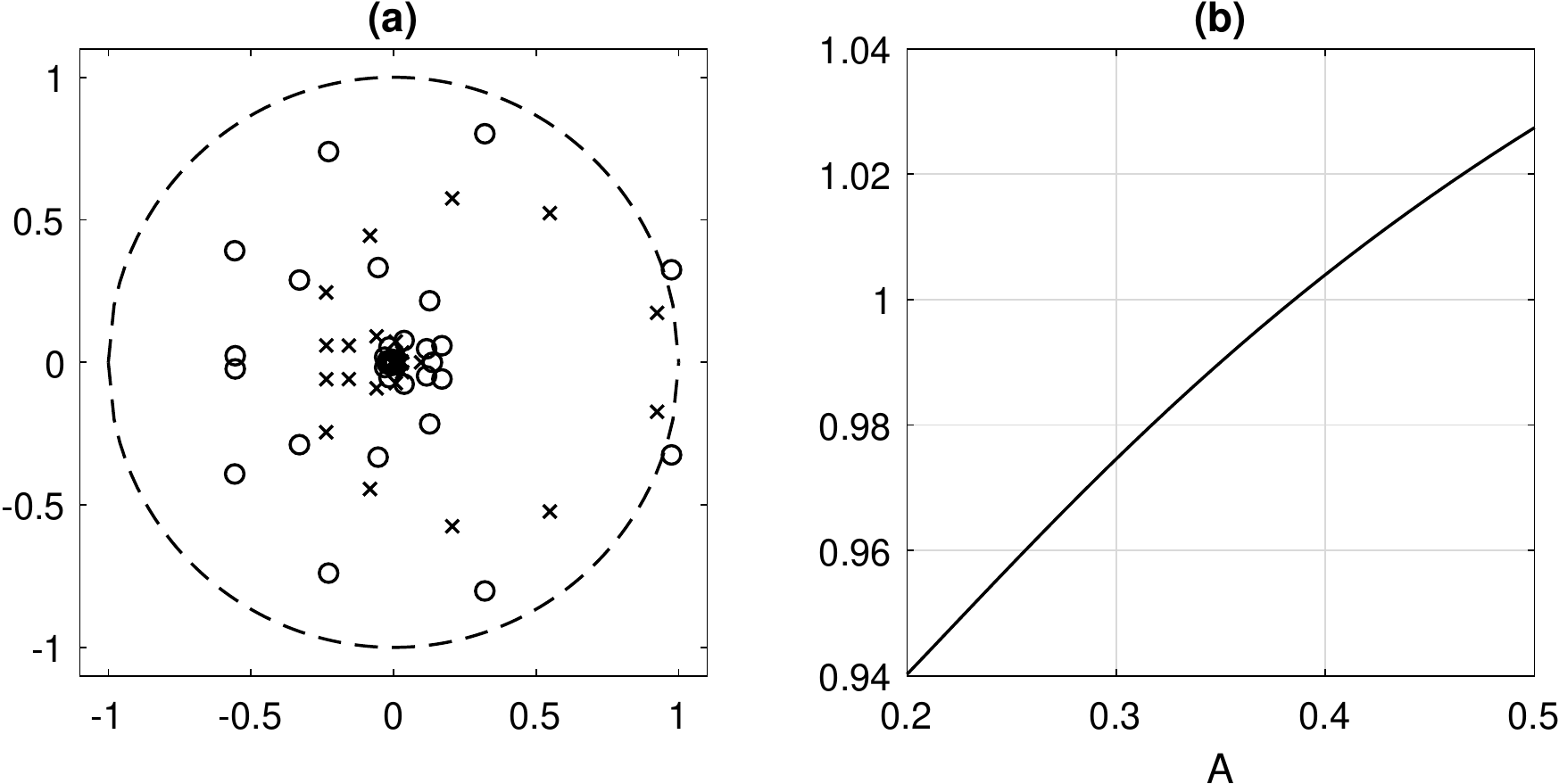}
\caption{Hopf bifurcation of a periodic solution of~\eqref{eq:dRdtFHN}-\eqref{eq:dPdtFHN} 
and~\eqref{eq:dUdt}-\eqref{eq:dVdt}. (a): Floquet multipliers of the periodic orbit at $A=0.2$
(crosses) and $A=0.5$ (circles). The unit circle is shown dashed. (b): maximum of the 
magnitude of the Floquet multipliers as a function of $A$.
Parameters: $\phi=\pi/2-0.1,\epsilon=0.1,a=0.5,D=10^{-3}$.
($\theta$ is discretised with 256 points.)}
\label{fig:HopfFHN}
\end{center}
\end{figure}

Performing numerical continuation of this periodic orbit in $A$ we find that it undergoes a Hopf
bifurcation as $A$ is increased, as shown in Fig.~\ref{fig:HopfFHN}. Numerical simulation
indicates that this is a supercritical bifurcation.
Decreasing $D$ decreases the value of $A$ at which the Hopf bifurcation occurs, suggesting that
the ``true'' bifurcation occurs at a lower value than that shown in  Fig.~\ref{fig:HopfFHN}.
Indeed, simulations of~\eqref{eq:dudtFHN}-\eqref{eq:dvdtFHN} with $N=5000$ show that
the Hopf bifurcation occurs at some $A\in(0.25,0.3)$.

(To  perform numerical continuation of the periodic orbit we define 
a Poincar{\'e} section at $V_2=0,\dot{V_2}>0$ and 
integrate~\eqref{eq:dRdtFHN}-\eqref{eq:dPdtFHN} 
and~\eqref{eq:dUdt}-\eqref{eq:dVdt} from an initial condition on this section
until the system hits the section for the first time. This defines a map in all other variables
from the section to itself, and a fixed point of this map is the periodic orbit of interest.
Linearising the map about the fixed point gives the  Floquet multipliers and hence
stability of the periodic orbit.)

\begin{figure}
\begin{center}
\includegraphics[width=13cm]{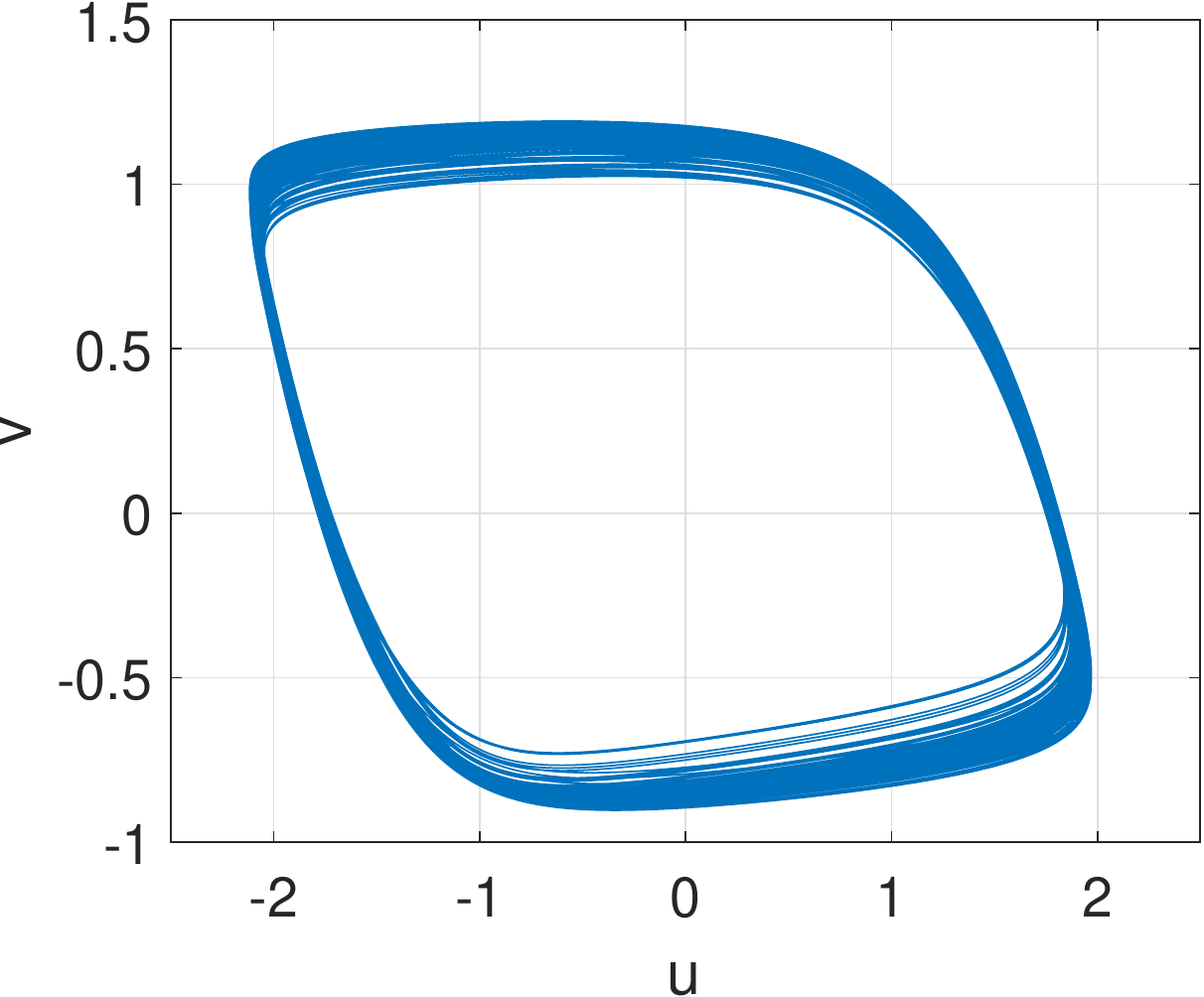}
\caption{Dynamics of~\eqref{eq:dudtFHNa}-\eqref{eq:dvdtFHNa} 
driven by~\eqref{eq:dRdtFHN}-\eqref{eq:dPdtFHN} 
and~\eqref{eq:dUdt}-\eqref{eq:dVdt}. 
Parameters: $A=0.2,\phi=\pi/2-0.1,\epsilon=0.1,a=0.5,D=10^{-3}$.
(Discretised with 256 points.)}
\label{fig:quasi}
\end{center}
\end{figure}

If we solve~\eqref{eq:dRdtFHN}-\eqref{eq:dPdtFHN} 
and~\eqref{eq:dUdt}-\eqref{eq:dVdt} with~\eqref{eq:PQa}-\eqref{eq:PQb} as drivers for
\begin{align}
   \epsilon\frac{du}{dt} & = u-u^3/3-v+\mu\left[b_{uu}\left(U_1-u\right)+b_{uv}\left(V_1-v\right)\right] \nonumber \\
  & +\nu\left[b_{uu}\left(U_2-u_i\right)+b_{uv}\left(V_2-v_i\right)\right] \label{eq:dudtFHNa}\\
   \frac{dv}{dt} & = u+a+\mu\left[b_{vu}\left(U_1-u\right)+b_{vv}\left(V_1-v\right)\right]+\nu\left[b_{vu}\left(U_2-u\right)+b_{vv}\left(V_2-v\right)\right] \label{eq:dvdtFHNa}
\end{align}
governing the dynamics of a single oscillator in the incoherent population, we find that $u$ and $v$
follow a stable quasiperiodic orbit, as shown in Fig.~\ref{fig:quasi}, 
with mean rotation frequency $\omega=2.1553$ while the synchronous
group (i.e.~$U_2$ and $V_2$) 
are periodic (as expected) with $\omega=2.0519$. These match quite well with the results
from simulating a finite network (Fig.~\ref{fig:FHN}) and differences could be due to
the finite $N$ used in Fig.~\ref{fig:FHN} and the non-zero value of $D$ needed to stabilise
the solutions of~\eqref{eq:dRdtFHN}-\eqref{eq:dPdtFHN} 
and~\eqref{eq:dUdt}-\eqref{eq:dVdt} ($D=10^{-3}$).

\subsubsection{Alternating chimera}
For a network formed from two populations, an alternating chimera may exist.
In this state neither population is pefectly synchronised and the level of synchrony
within each population varies periodically, but in antiphase to that of the other 
population~\cite{hausch15}. One way that such a state can form is that under parameter variation,
two coexisting ``breathing'' chimeras, in which one population 
is synchronised and the other is not (which
are mapped to one another under relabelling of the populations),
merge in a gluing bifurcation, resulting in an attractor which is invariant
under relabelling of the populations~\cite{lai12}.

Such a state occurs in~\eqref{eq:dudtFHN}-\eqref{eq:dvdtFHN} for $A=0.6$, i.e.~after the
Hopf bifurcation.
Since oscillators in both populations now lie on (different) closed curves, we can write the
dynamics for each curve. 
The equations governing the system are
\begin{align}
   \frac{\p R_1}{\p t}(\theta,t) & = F(R_1,\theta,U_1,V_1,U_2,V_2) -G(R_1,\theta,U_1,V_1,U_2,V_2)\frac{\p R_1}{\p \theta}+D\frac{\p^2}{\p \theta^2}R_1(\theta,t)  \label{eq:dR1dt} \\
   \frac{\p P_1}{\p t}(\theta,t) & = -\frac{\p}{\p \theta}\left[P_1(\theta,t)G(R_1,\theta,U_1,V_1,U_2,V_2)\right] +D\frac{\p^2}{\p \theta^2}P_1(\theta,t) 
\end{align}
and
\begin{align}
   \frac{\p R_2}{\p t}(\theta,t) & = F(R_2,\theta,U_2,V_2,U_1,V_1) -G(R_2,\theta,U_2,V_2,U_1,V_1)\frac{\p R_2}{\p \theta}+D\frac{\p^2}{\p \theta^2}R_2(\theta,t)  \\
   \frac{\p P_2}{\p t}(\theta,t) & = -\frac{\p}{\p \theta}\left[P_2(\theta,t)G(R_2,\theta,U_2,V_2,U_1,V_1)\right] +D\frac{\p^2 P_2(\theta,t)}{\p \theta^2}  \label{eq:dP2dt}
\end{align}
where
\be
   U_j=\int_0^{2\pi}P_j(\theta,t)R_j(\theta,t)\cos{\theta}d\theta
\ee
and
\be 
  V_j=\int_0^{2\pi}P_j(\theta,t)R_j(\theta,t)\sin{\theta}d\theta
\ee

\begin{figure}
\begin{center}
\includegraphics[width=13cm]{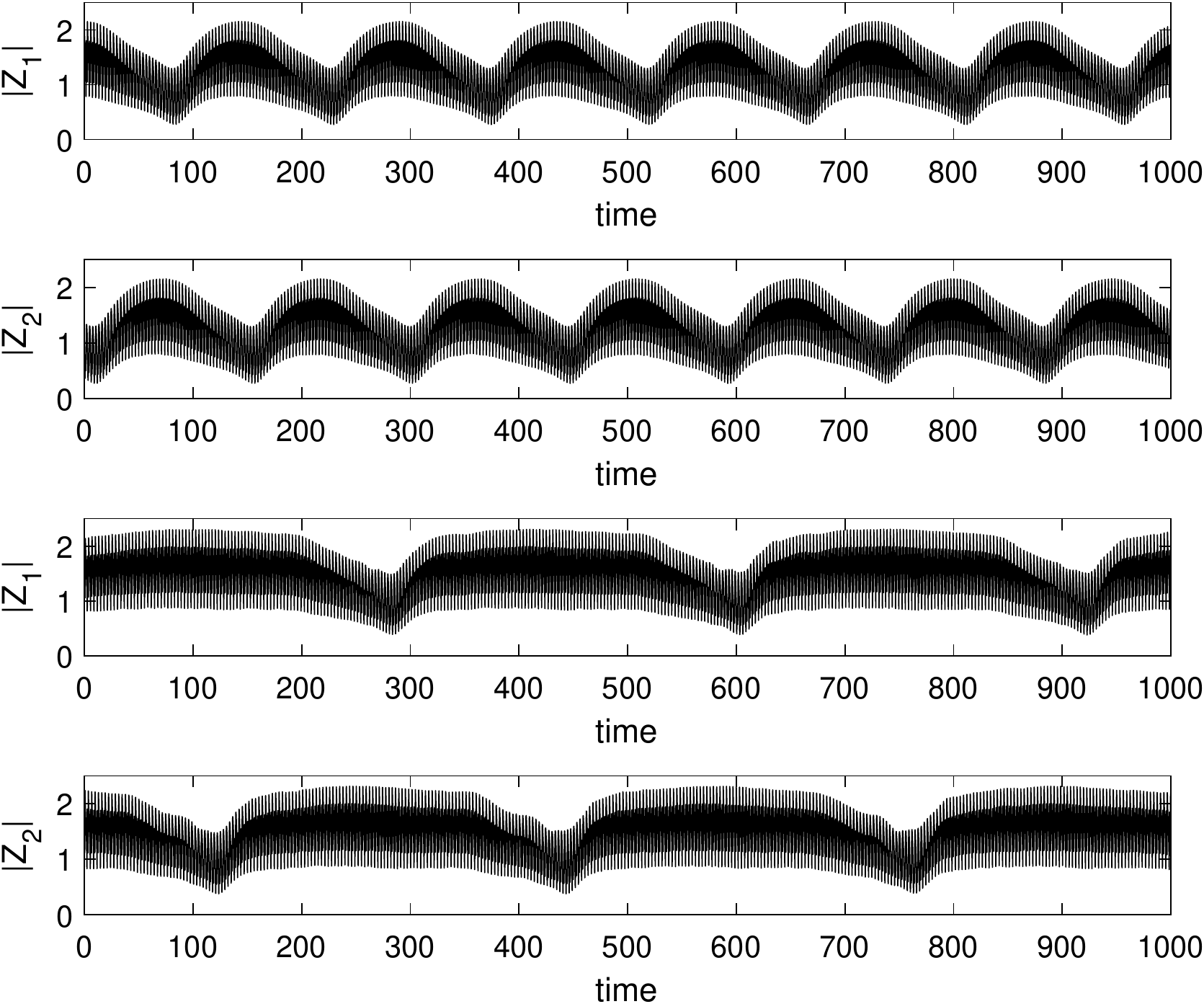}
\caption{An alternating chimera state. Top two panels: magnitudes of the order parameters
defined for~\eqref{eq:dR1dt}-\eqref{eq:dP2dt} (discretised in $\theta$
using 256 points, $D=10^{-4}$.). Bottom two panels:
magnitudes of the order parameters
defined for~\eqref{eq:dudtFHN}-\eqref{eq:dvdtFHN} (with $N=500$).
Parameters: $A=0.6,\phi=\pi/2-0.1,\epsilon=0.1,a=0.5$.}
\label{fig:breath}
\end{center}
\end{figure}

To quantify the behaviour we 
define order parameters $Z_j=U_j+iV_j$ and plot the magnitude of both of these in the top
two panels of Fig.~\ref{fig:breath}. To compare with the behaviour 
of~\eqref{eq:dudtFHN}-\eqref{eq:dvdtFHN} we define
\be
   Z_1=\frac{1}{N}\sum_{k=1}^N u_k+\frac{i}{N}\sum_{k=1}^N v_k
\ee
and
\be
   Z_2=\frac{1}{N}\sum_{k=1}^N u_{N+k}+\frac{i}{N}\sum_{k=1}^N v_{N+k}
\ee
and plot their magnitudes in the bottom two panels of Fig.~\ref{fig:breath} for $N=500$.
We see alternations, as expected, and the range of values and the
form of oscillations is correct.
The main difference between the two systems
is the timescale of alternation. This is probably due to the finite size of the
population in~\eqref{eq:dudtFHN}-\eqref{eq:dvdtFHN},
the non-zero value of $D$ used in~\eqref{eq:dR1dt}-\eqref{eq:dP2dt}, 
and presumed closeness to a gluing bifurcation, in which two symmetrically related breathing
chimeras merge to form the alternating chimera, as in~\cite{lai12}. Since this bifurcation
involves a quasiperiodic orbit approaching a saddle periodic orbit, we expect that the time
it spends near the saddle orbit, and thus the period of the slow oscillations seen in
Fig.~\ref{fig:breath}, to be quite sensitive to the differences between the two systems being
studied here.

\subsection{Delay}
\label{sec:del}

Chimeras have been studied in a number of systems with delays~\cite{omemai08,setsen08,lai09}.
In this section we consider the system
\begin{align}
   \frac{dX_j(t)}{dt} & = i\omega X_j(t)+\gamma(1-|X_j(t)|^2)X_j(t) +\frac{\mu}{N}\sum_{k=1}^N X_k(t-\tau_1)+\frac{\nu}{N}\sum_{k=1}^N X_{N+k}(t-\tau_2) \label{eq:dXdtdelA}
\end{align}
for $j=1,\dots N$ and
\begin{align}
   \frac{dX_j(t)}{dt} & = i\omega X_j(t)+\gamma(1-|X_j(t)|^2)X_j(t) +\frac{\mu}{N}\sum_{k=1}^N X_{N+k}(t-\tau_1)+\frac{\nu}{N}\sum_{k=1}^N X_{k}(t-\tau_2) \label{eq:dXdtdelB}
\end{align}
for $j=N+1,\dots 2N$, where each $X_j\in\mathbb{C}$, 
i.e.~two populations of Stuart-Landau oscillators with coupling within a
population of strength $\mu$, delayed by $\tau_1$,
and coupling between populations of strength $\nu$, delayed by $\tau_2$. We find that
there is a chimera
for parameters $\omega=3,\gamma=10, \mu=0.36,\nu=0.04,\tau_1=0.6,\tau_2=0.4$, in a system
with $N=100$ (not shown). The asynchronous group has an average angular frequency of $2.9871$, 
and the synchronised group has angular frequency $2.6348$.

Suppose population two is synchronised. Then 
its dynamics is described by
\be
   \frac{dX(t)}{dt}  = i\omega X(t)+\gamma(1-|X(t)|^2)X(t)+\mu X(t-\tau_1)+\nu Z(t-\tau_2) \label{eq:dXdtdel}
\ee
where
\be
   Z(t)=\frac{1}{N}\sum_{k=1}^N X_{k}(t)
\ee
In population one we have
\begin{align}
   \frac{dX_j(t)}{dt} & = i\omega X_j(t)+\gamma(1-|X_j(t)|^2)X_j(t)+\mu Z(t-\tau_1)+\nu X(t-\tau_2) 
\end{align}
for $j=1,\dots N$. Writing $X_j=r_je^{i\theta_j}$ we have
\begin{align}
   \frac{dr_j}{dt} & = \gamma[1-r_j^2(t)]r_j(t) +\mbox{Re}\left\{[\mu Z(t-\tau_1)+\nu X(t-\tau_2)]e^{-i\theta_j(t)}\right\} \nonumber \\
   & \equiv F[r_j(t),\theta_j(t),Z(t-\tau_1),X(t-\tau_2)] \\
   \frac{d\theta_j}{dt} & = \omega+\mbox{Im}\left\{[\mu Z(t-\tau_1)+\nu X(t-\tau_2)]e^{-i\theta_j}(t)\right\}/r_j(t) \nonumber \\
   & \equiv G[r_j(t),\theta_j(t),Z(t-\tau_1),X(t-\tau_2)]
\end{align}

Thus we consider the dynamical system
\begin{align}
   \frac{\p R}{\p t}(\theta,t) & = F[R(\theta,t),\theta,Z(t-\tau_1),X(t-\tau_2)] \nonumber \\
   & -G[R(\theta,t),\theta,Z(t-\tau_1),X(t-\tau_2)]\frac{\p R}{\p \theta}(\theta,t) \\ 
   \frac{\p P}{\p t}(\theta,t) & = -\frac{\p}{\p \theta}\left\{P(\theta,t)G[R(\theta,t),\theta,Z(t-\tau_1),X(t-\tau_2)]\right\} +D\frac{\p^2}{\p \theta^2}P(\theta,t) 
\end{align}
together with~\eqref{eq:dXdtdel} where
\be
   Z(t)=\int_0^{2\pi} P(\theta,t)R(\theta,t)e^{i\theta}d\theta
\ee
We set the level of diffusion to be $D=10^{-6}$. 

\begin{figure}
\begin{center}
\includegraphics[width=13cm]{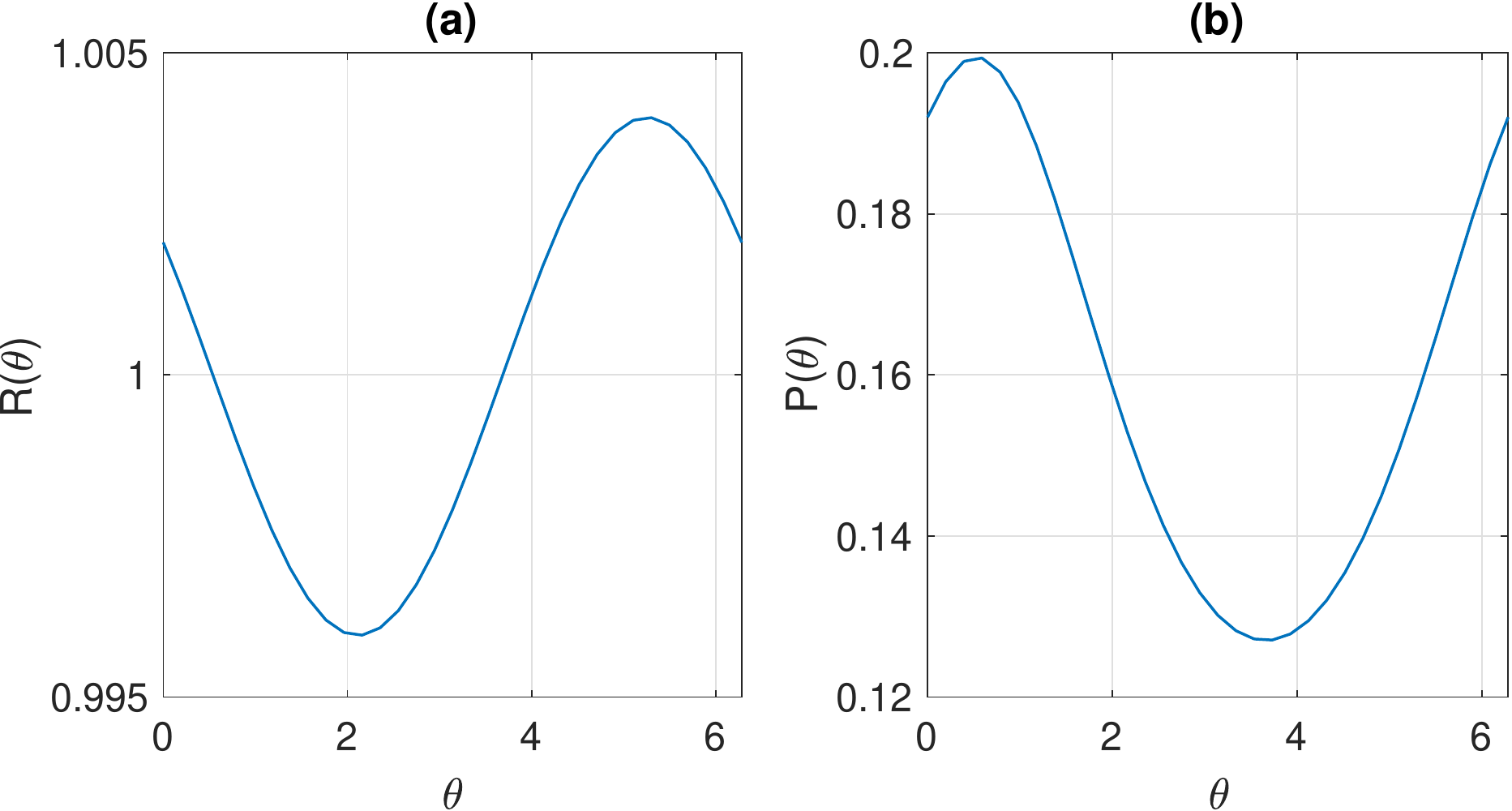}
\caption{Steady state of~\eqref{eq:dRtdt}-\eqref{eq:dPtdt} and~\eqref{eq:dXtdtdel}.
Parameters: $\omega=3,\gamma=10, \mu=0.36,\nu=0.04,\tau_1=0.6,\tau_2=0.4, D=10^{-6}$.}
\label{fig:snapdel}
\end{center}
\end{figure}

This system is invariant under rotation in the complex plane of each $X_j$ by the same angle
so we can go to a rotating coordinate frame in which the chimera is stationary. In this frame,
defining $\tilde{X}(t)=X(t)e^{-i\Omega t}$ and $\tilde{Z}(t)=Z(t)e^{-i\Omega t}$ we find that
$\tilde{X}$ satisfies
\begin{align}
   \frac{d\tilde{X}(t)}{dt} & = i(\omega-\Omega) \tilde{X}(t)+\gamma(1-|\tilde{X}(t)|^2)\tilde{X}(t) +\mu \tilde{X}(t-\tau_1)e^{-i\Omega\tau_1}+\nu \tilde{Z}(t-\tau_2)e^{-i\Omega \tau_2} \label{eq:dXtdtdel}
\end{align}
where $\Omega$ is the speed of rotation. Note that moving to a rotating frame causes
effective phase shifts in $\tilde{X}$ and $\tilde{Z}$.
Writing $\tilde{X}_j(t)=X_j(t)e^{-i\Omega t}=\tilde{r}_j(t)e^{i\tilde{\theta}_j(t)}$ we find that
in population one,
\begin{align}
   \frac{d\tilde{r}_j}{dt} & = \gamma[1-\tilde{r}_j^2(t)]\tilde{r}_j(t) +\mbox{Re}\left\{[\mu \tilde{Z}(t-\tau_1)e^{-i\Omega\tau_1}+\nu \tilde{X}(t-\tau_2)e^{-i\Omega\tau_2}]e^{-i\tilde{\theta}_j(t)}\right\} \nonumber \\
   & \equiv \tilde{F}[\tilde{r}_j(t),\tilde{\theta}_j(t),\tilde{Z}(t-\tau_1),\tilde{X}(t-\tau_2)] \\
   \frac{d\tilde{\theta}_j}{dt} & = \omega-\Omega +\mbox{Im}\left\{[\mu \tilde{Z}(t-\tau_1)e^{-i\Omega\tau_1}+\nu \tilde{X}(t-\tau_2)e^{-i\Omega\tau_2}]e^{-i\tilde{\theta}_j(t)}\right\}/\tilde{r}_j(t) \nonumber \\
   & \equiv \tilde{G}[\tilde{r}_j(t),\tilde{\theta}_j(t),\tilde{Z}(t-\tau_1),\tilde{X}(t-\tau_2)]
\end{align}
We are thus interested in steady states of
\begin{align}
   \frac{\p R}{\p t}(\theta,t) & = \tilde{F}[R(\theta,t),\theta,\tilde{Z}(t-\tau_1),\tilde{X}(t-\tau_2)] \nonumber \\
   & -\tilde{G}[R(\theta,t),\theta,\tilde{Z}(t-\tau_1),\tilde{X}(t-\tau_2)]\frac{\p R}{\p \theta}(\theta,t) \label{eq:dRtdt} \\ 
   \frac{\p P}{\p t}(\theta,t) & = -\frac{\p}{\p \theta}\left\{P(\theta,t)\tilde{G}[R(\theta,t),\theta,\tilde{Z}(t-\tau_1),\tilde{X}(t-\tau_2)]\right\}+D\frac{\p^2}{\p \theta^2}P(\theta,t) \label{eq:dPtdt} 
\end{align}
along with~\eqref{eq:dXtdtdel} where
\be
   \tilde{Z}(t)=\int_0^{2\pi} P(\theta,t)R(\theta,t)e^{i\theta}d\theta
\ee
Such a steady state is shown in Fig.~\ref{fig:snapdel}, where $\Omega=2.6375$.
(Matlab's dde23 routine was used for time integration.)
Numerical study of delay differential equations is significantly more difficult than
that of non-delayed equations, so
we discretise $\theta$ in only 32 equally spaced points. As can be seen in Fig.~\ref{fig:snapdel}
the solutions are quite smooth functions of $\theta$, and spatial derivatives are evaluated
spectrally.

Having found the steady state of~\eqref{eq:dRtdt}-\eqref{eq:dPtdt} and~\eqref{eq:dXtdtdel} we can 
numerically integrate the ODEs
\be
   \frac{dr}{dt}  = \tilde{F}[r,\theta,\tilde{Z},\tilde{X}]; \qquad 
   \frac{d\theta}{dt}  = \tilde{G}[r,\theta,\tilde{Z},\tilde{X}] \label{eq:ODEdel}
\ee
where $\tilde{Z}$ and $\tilde{X}$ no longer depend on time, in order
to find the period of an oscillator in the incoherent group relative to the frequency of the
locked group ($\Omega$). For the parameters used here,~\eqref{eq:ODEdel} has a stable
periodic orbit with angular frequency $\sim 0.35338$, showing that the coherent and incoherent groups
do have different average frequencies, as expected for a chimera state. Adding this frequency
to the measured value of $\Omega$
we obtain $2.9909$, in very good agreement with the measured angular frequency
from the finite simulation ($2.9871$).

\begin{figure}
\begin{center}
\includegraphics[width=13cm]{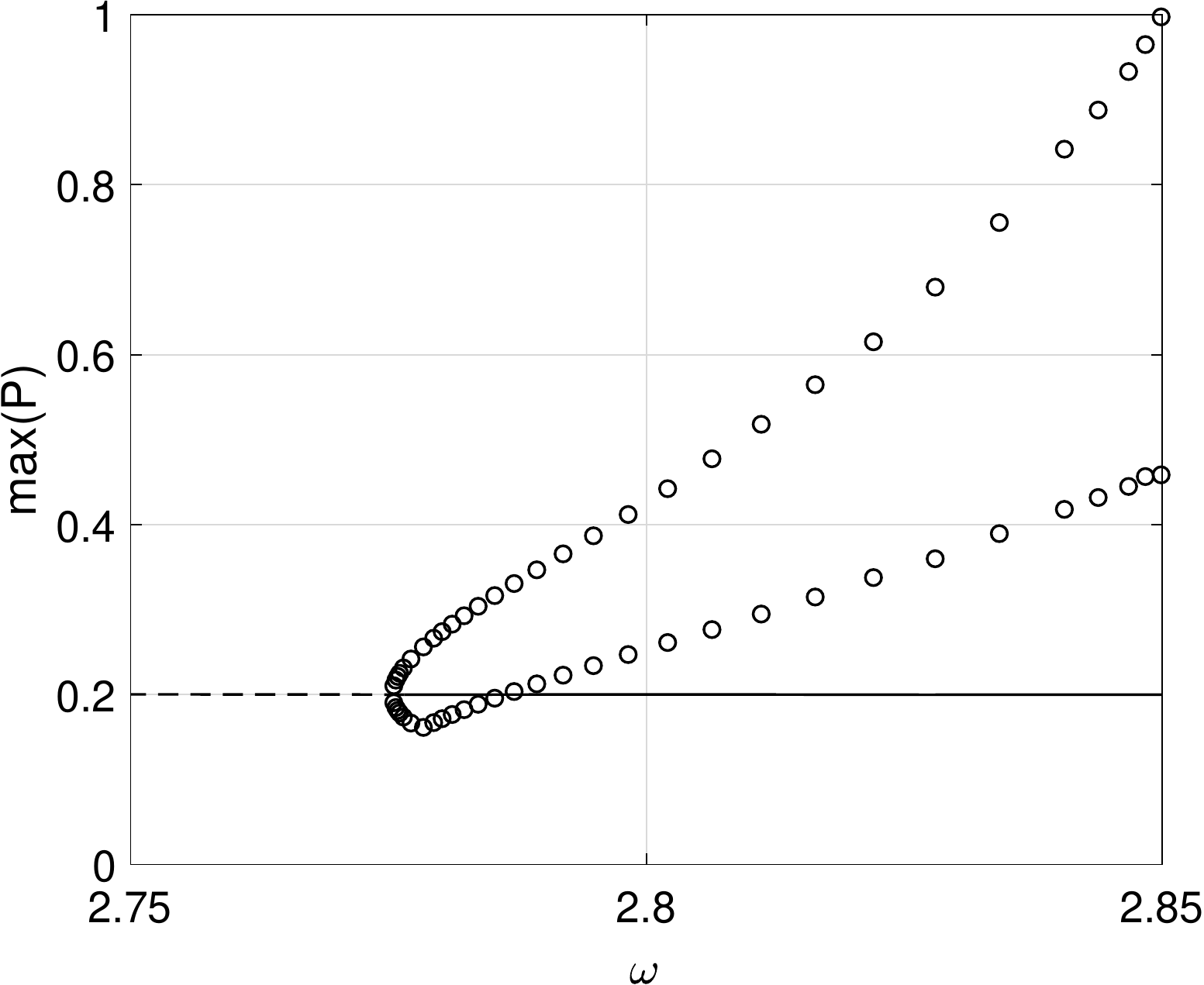}
\caption{Lines: maximum of $P(\theta)$ at steady states of~\eqref{eq:dRtdt}-\eqref{eq:dPtdt} 
and~\eqref{eq:dXtdtdel} (solid: stable; dashed: unstable). Open circles: unstable periodic
orbit created in a subcritical Hopf bifurcation (see text).
Parameters: $\gamma=10, \mu=0.36,\nu=0.04,\tau_1=0.6,\tau_2=0.4, D=10^{-6}$.
(Discretised with 32 points.)}
\label{fig:Hopf}
\end{center}
\end{figure}

We can following the steady state shown in Fig.~\ref{fig:snapdel} 
as $\omega$ is decreased using the software 
DDE-BIFTOOL~\cite{ddebif}. Doing so we find that it becomes unstable through
a subcritical Hopf bifurcation, as shown in Fig.~\ref{fig:Hopf}. 
We can also follow the unstable periodic orbit created in this bifurcation
as $\omega$ is varied.
To represent the unstable
periodic orbit we track the maximum over $\theta$ of $P(\theta,t)$, and then show the
maximum and minimum values over one period of this, with open circles.

\begin{figure}
\begin{center}
\includegraphics[width=13cm]{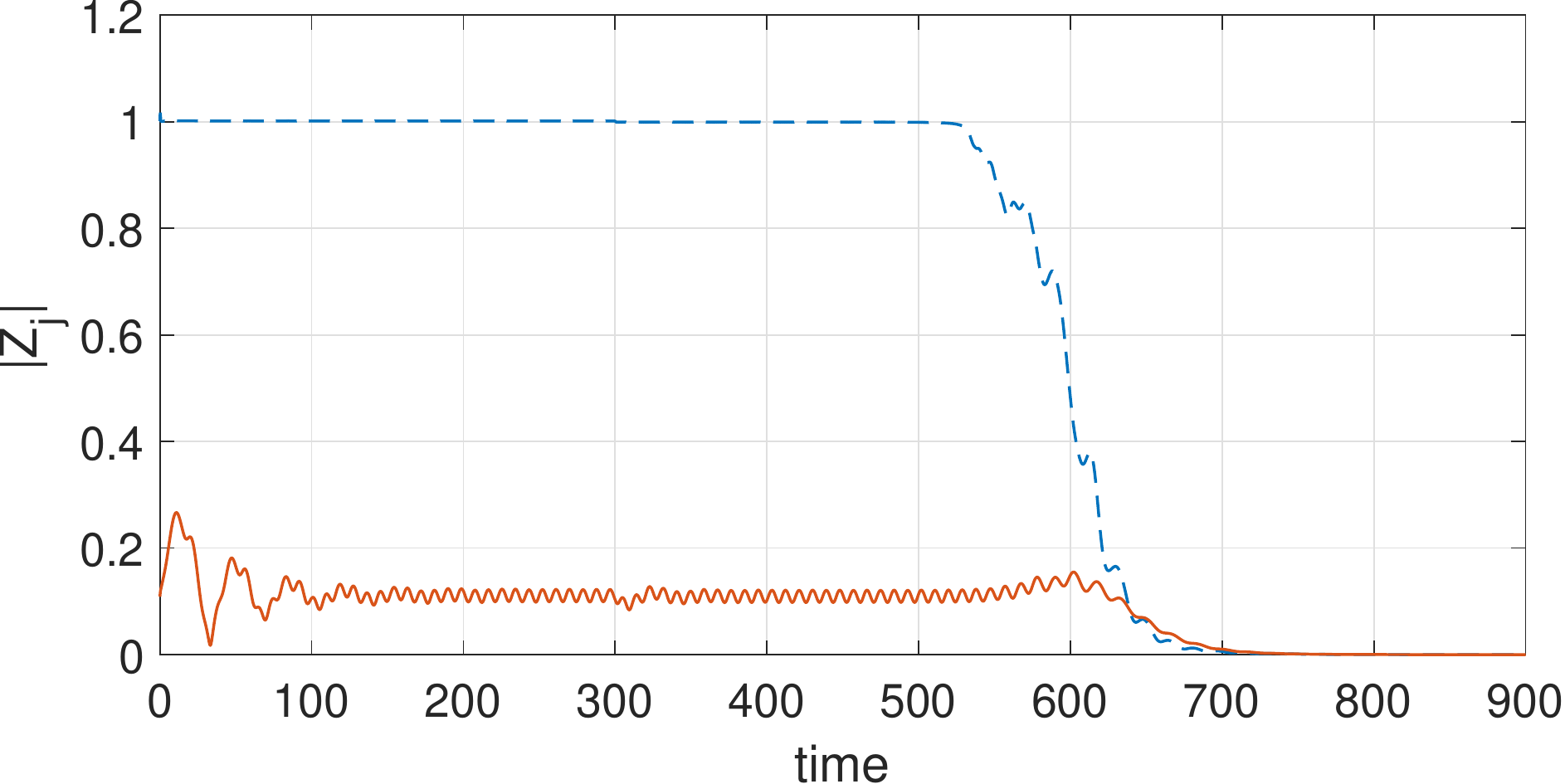}
\caption{Solution of~\eqref{eq:dXdtdelA}-\eqref{eq:dXdtdelB}. $Z_1(t)\equiv N^{-1}\sum_{k=1}^N 
X_k(t)$ and $Z_2(t)\equiv N^{-1}\sum_{k=1}^N X_{N+k}(t)$.
$\omega$ is increased from 3 to 3.2 at $t=300$.
Parameters: $\gamma=10, \mu=0.36,\nu=0.04,\tau_1=0.6,\tau_2=0.4, N=100$.
}
\label{fig:destab}
\end{center}
\end{figure}

Increasing $\omega$ in the discrete network~\eqref{eq:dXdtdelA}-\eqref{eq:dXdtdelB} to $\omega=3.2$
destabilises the chimera state, and the system moves to a state where both populations
 are incoherent, as shown in Fig.~\ref{fig:destab}. Such an instability cannot be detected
using the approach presented above, which assumes that one population is perfectly
synchronised.

\section{Discussion}
\label{sec:disc}

We have used the results of~\cite{clupol18} to study the dynamics of chimera states
in networks formed from two populations of identical oscillators, with different strengths
of coupling both within and between populations. We studied four different types of
oscillators. In Sec.~\ref{sec:SL} we revisited the system of Stuart-Landau
oscillators studied in~\cite{lai10} and put results that were inferred in that paper on a 
solid footing. In Sec.~\ref{sec:inertia} we consider Kuramoto oscillators with
inertia, previously studied in~\cite{boukan14,olm15}. We showed that stable
stationary chimeras do not exist is such systems, at least for an infinite number
of oscillators and for the parameter values previously considered. In Sec.~\ref{sec:FHN}
we considered FitzHugh-Nagumo oscillators whose oscillations are highly nonlinear.
This system is unlike the three others studied, as the oscillators are not invariant
under a phase shift, and thus the chimera state of interest is actually a periodic orbit
rather than a fixed point in a rotating coordinate frame. Lastly (Sec.~\ref{sec:del})
we considered Stuart-Landau oscillators with delayed coupling. We have provided rigorous
numerical results on the existence and stability of chimeras in these networks, in contrast
to the many presentations showing results of only numerical simulations of finite networks
of oscillators.

\begin{figure}
\begin{center}
\includegraphics[width=13cm]{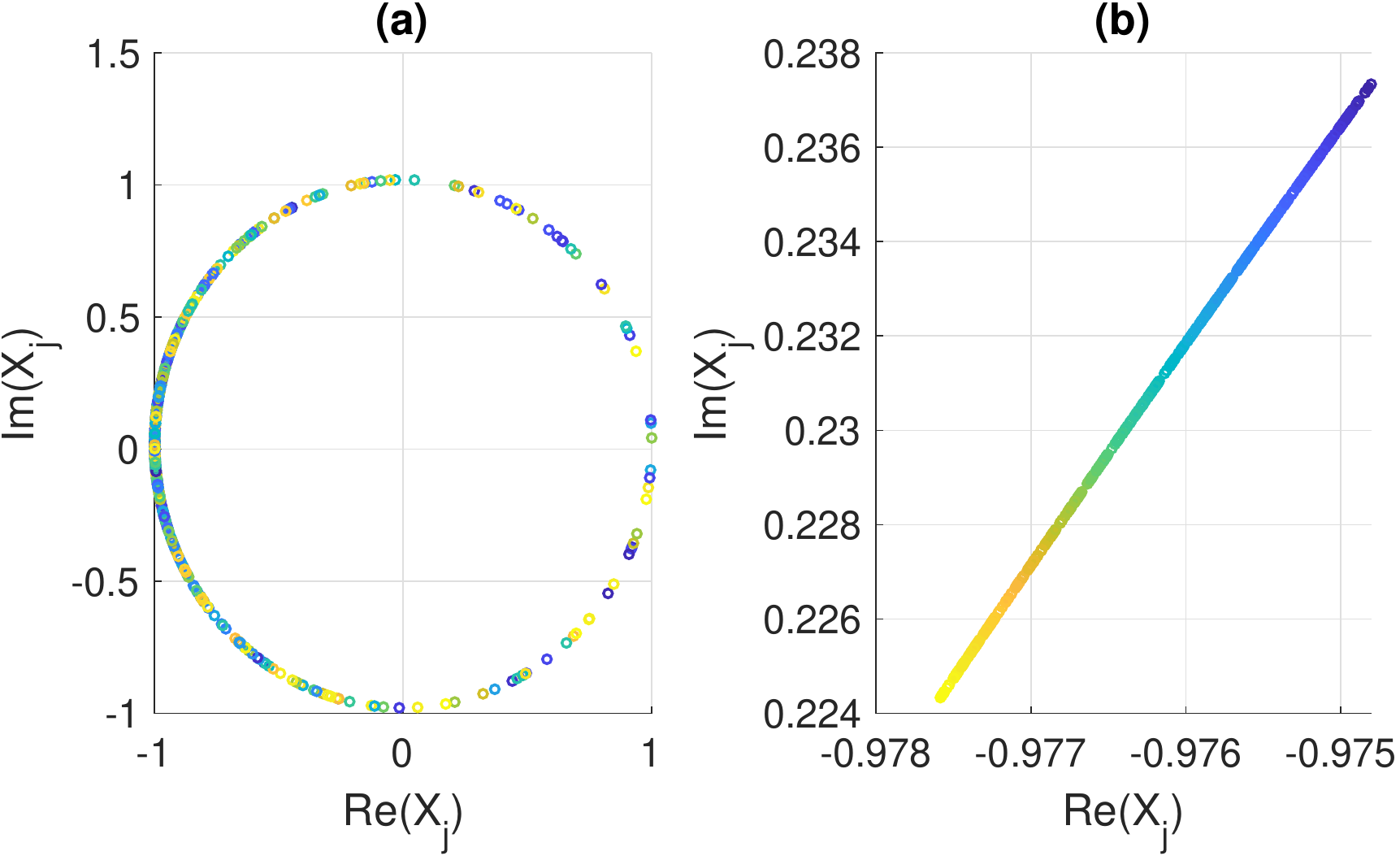}
\caption{Snapshot of a solution of~\eqref{eq:dXdt1}-\eqref{eq:dXdt2} where for each
oscillator, $\omega$ was randomly chosen from a uniform distribution on $[0,0.002]$.
(a): asynchronous population; (b): nearly synchronous population. Note the different
scales. Colour indicates the
value of each $\omega$. 
Parameters: $\epsilon=0.05,\beta=0.08,A=0.2,\delta=-0.1, N=500$.
}
\label{fig:het}
\end{center}
\end{figure}

Regarding future work,
all of the results presented here consider identical oscillators. However, at least
for sinusoidally-coupled phase oscillators it is known that systems of identical
oscillators have non-generic properties such as a large number of conserved
quantities~\cite{watstr94}, and making them heterogeneous removes this 
degeneracy~\cite{lai09a,ottant08}. To investigate this 
we numerically integrated~\eqref{eq:dXdt1}-\eqref{eq:dXdt2},
but having made the system heterogeneous by choosing the value of $\omega$ for each
oscillator randomly and independently from a uniform distribution. A snapshot
of the solution is shown in Fig.~\ref{fig:het}, where the oscillators are coloured by their
$\omega$ value. This state can still be regarded as a chimera, as it is a small perturbation
from the chimera that exists for identical oscillators. For both populations, the oscillators
lie on a smooth curve. However, for the asynchronous population there seems to be no
correspondence between the value of $\omega$ and an oscillator's position on the curve,
while in the nearly synchronous group the oscillators are clearly ordered by the value of
$\omega$. It may be possible to derive a theory to cover this type of solution. 

It would also be of interest to develop a theory for oscillators described by more
than two variables, assuming that the incoherent oscillators still lie on a closed
curve in phase space. 

While we have considered abstract networks of oscillators, the modelling of neurons
or groups of neurons by oscillators is common~\cite{ashcoo16}. A network of two populations,
as studied here, naturally arises when modelling the dynamics of competition between
two competing percepts, for example in binocular rivalry~\cite{laicho02}. The techniques
presented here may be useful in further understanding the dynamics of such networks.

\end{document}